\documentclass[fleqn, 12pt]{article}
\usepackage{graphics,times}
\begin{document}
\title{Gravitation in the fractal $D=2$ inertial universe: \\ New phenomenology in
spiral discs and a \\ theoretical basis for MOND}
\author{D. F. Roscoe \\
School of Mathematics, \\
 Sheffield University, Sheffield, S3 7RH, UK. \\
Email: D.Roscoe@ac.shef.uk \\
Tel: 0114-2223791, Fax: 0114-2824292}
\maketitle
\begin{abstract}
A particular interpretation of Mach's Principle led us to consider
if it was possible to have a globally inertial universe that was irreducibly
associated with a non-trivial global matter distribution, Roscoe \cite{RoscoeC}.
This question received a positive answer, subject to the condition
that the global matter distribution is necessarily fractal, $D=2$.
The purpose of the present paper is to show how general gravitational
processes arise in this universe. We begin by showing how classical
Newtonian gravitational processes arise from point-source perturbations
of this $D=2$ inertial background. We are then able to use the insights
gained from this initial analysis to arrive at a general theory for
arbitrary material distributions. We illustrate the process by using
it to model an idealized spiral galaxy. One particular subclass of solutions,
corresponding to {\it logarithmic} spirals, has
already been extensively tested (Roscoe \cite{RoscoeA}, \cite{RoscoeD}),
and shown to resolve dynamical data over large samples of optical
rotation curves (ORCs) with a very high degree of statistical precision.

Whilst the primary purpose of the data analysis of Roscoe \cite{RoscoeA}
was to test the predictions of the logarithmic spiral theory, it led directly
to the discovery of a major new phenomenology in spiral discs - that
of \textit{discrete dynamical classes} - initially reported in Roscoe
\cite{RoscoeB} and comprehensively confirmed in Roscoe \cite{RoscoeD}
over four large independent samples of ORCs. In this paper, we analyse the
theory more comprehensively, and show  how the discrete dynamical classes
phenomenology has a ready explanation in terms of an algebraic
consistency condition which must necessarily be satisfied.

Of
equal significance, we apply the theory with complete success to the detailed
modelling of a sample of eight Low Surface Brightness spirals (LSBs) which,
hitherto, have been succesfully modelled only by the MOND algorithm (Modified
Newtonian Dynamics, Milgrom \cite{Milgrom1983a}, \cite{Milgrom1983b},
\cite{Milgrom1983c}). The CDM models have failed comprehensively when applied to
LSBs. We are able to conclude that the essence of the MOND algorithm must be
contained within the presented theory.
\hfill{}\break \hfill{}\break 
\leftline{\bf{Mach -- Spiral Galaxy -- Rotation Curve -- MOND -- Gravitation}}
\end{abstract} \section{Introduction}
\subsection{Review of Preliminary Work:} \label{Sec1}
General Relativity can be considered as a theory of what happens to
predetermined clocks and rods in the presence of material
systems (material here is understood in its widest sense). By contrast,
the theory being discussed here  can be considered as a
primitive, but fundamental, theory of how clocks and rods arise in the first
place within material systems.

The preliminary work (Roscoe \cite{RoscoeC}) was driven by the
idea that it is impossible to conceive of physical (metric) space in the
\emph{absence} of material systems, and that it is similarly impossible to
conceive of physical time in the absence of {\emph {process}} within these
material systems. Following upon this, we took the point of view that
any {{ fundamental}} theory of space \& time must then necessarily have the
property that, within it, it is {{ impossible}} to talk about metrical
space \& physical time in the absence of any material system.   In effect, this
is the view that notions of material systems are logically prior to
notions of clocks \& rods, and that these latter notions are somehow
\emph{projected} out of prior relations which exists between the individual
elements which make up the former.

We began by noting that the most simple form of
space \& time we can conceive is that of inertial space \& time - that is, a
space \& time within which the relations between
clocks and rods are fixed. And then, keeping to the spirit of the basic idea, we
posed the questions:
\begin{it}
\begin{itemize}
\item
Can a globally inertial space \& time be associated with a non-trivial global
matter distribution ?
\item And, if so, what are the general properties of this distribution ?
\end{itemize}
\end{it}
These questions were addressed within the context of an extremely simple model
universe populated by particles which possessed only the property of
\emph{enumerability}, and within which there were no predetermined ideas of
clocks \& rods. It was required that these concepts should emerge from the
general analysis. To simplify the initial development, it was originally assumed
that the model universe was stationary. It transpired that this assumption was
equivalent to choosing a one-clock quasi-classical model, whilst relaxing the
assumption was equivalent to choosing a two-clock relativistic model. The work
of this paper is based upon the one-clock quasi-classical development.

The original questions were then answered as follows: a globally inertial space \&
time  can be associated with a non-trivial matter distribution, and this
distribution is necessarily fractal $D=2$.
However, it transpires that the particles in this matter distribution cannot
reasonably be identified with ordinary matter since (in crude terms) the
particles all appear to be in states of randomly directed uniform motion,
but with identical speeds of magnitude $v_0$. In other words, the
distribution has some of the attributes of a quasi-photon gas and, for this
reason, we interpreted it as a rudimentary model of a material vacuum.  A closer
investigation then shows that, in fact, whilst $v_0$ has the dimensions of
speed, it is more properly interpreted as a conversion factor between length
scales and temporal scales - in this sense, $v_0$ is more like Bondi's
interpretation of $c$.  It follows from this that the material vacuum itself
appears to have the role of arbitrating between length scales and temporal
scales.

We then noted that, if ordinary matter could, somehow, condense out of this
$D=2$ material vacuum then we would have a universe of ordinary matter which is
in close accord with what is actually observed in
galaxy counts out to medium distances - that is, on these medium scales, the
material distribution in the form of galaxies has vanishingly small
accelerations and is distributed in a quasi-fractal form with $D \approx 2$.
\subsection{The Emergence of Gravitation}
Once one has a particular conception of {\lq inertial space \& time'}
(for example, that of Newtonian theory, or that of Einstein's
special relativity), a theory of gravitation effectively follows as a
perturbation of that particular inertial space \& time.

For example, in the Newtonian context, such perturbations
are interpreted as the introduction of {\it gravitational forces} into a
previously force-free environment whilst, in the Einsteinien case, they are
interpreted as the introduction of {\it curvature} into a flat spacetime
manifold. In the present case, they are literally perturbations of the $D=2$
distribution of material in the rudimentary model vacuum. Since the $D=2$ distribution
is irreducibly associated with clocks \& rods in a fixed relationship with each
other, it follows that any perturbation from $D=2$ will necessarily entail
distortions of the clocks \& rods relationship, and will therefore give rise to
what are conventionally called {\emph gravitational processes}.
\subsection{Qualitative Gravitational Mechanisms}
Newton introduced the idea of {\bf \emph{force}} into discourse
about the world through his mechanics, and his Law of Gravitation can be viewed
as a recipe quantifying the amount of {\bf \emph{force}} acting between two
massive bodies. However, Newton famously said that he had no idea what
constituted the fundamental essence of this {\bf \emph{gravitational force}} and
today, the whole theory is simply accepted as an enormously successful and
useful means of describing the phenomenology.

Exactly the same can be said of General Relativity: that is, gravitational
processes are said to arise in this theory when the flat spacetime manifold
becomes curved by the presence of mass. The field equations can be viewed simply
as the recipe which quantifies the amount of curvature created by a given
distribution of mass. But, as with Newtonian theory, the {\bf \emph{mechanism}}
by which this is achieved is absent. So, at one level, General Relativity can be
considered as simply an alternative means of describing the phenomenology -
albeit one which applies in far more extreme circumstances.

The case of the present theory is different: the search for a {\bf
\emph{gravitational mechanism}} was emphatically not part of our original
thinking and, in the main body of this paper, we pay no attention at all to the
likely nature of any such mechanism. We simply perform a formal perturbation
analysis of the $D=2$ inertial universe, and the analysis is played out
geometrically on a curved manifold using largely familiar techniques.
However, regardless of this, it is impossible not to realize that a genuine
mechanism has automatically presented itself. Specifically, we have
mentioned that the material in the idealized $D=2$ inertial universe behaves
like a quasi-photon gas, and that gravitational processes arise when this $D=2$
distribution is perturbed - in particular, point-wise spherically symmetric
perturbations are interpreted as being due to the presence of conventional point
masses. One is then immediately forced to conclude that the point-mass
perturbs by acting as either a {\bf \emph{source}} or a {\bf \emph{sink}} of
quasi-photon gas.  Suppose it is a sink, and that its mass simply quantifies the
amount  of absorption going on; then, when two such absorbing masses are placed
in the $D=2$ universe, they will partly shade each other from the global
quasi-photon flux,  and will therefore experience a net external pressure acting
to push them together. That is, a {\bf \emph{gravitational force}} arises. We
merely note that such explicit mechanisms have been proposed before: the
difference here is that the {\bf \emph{absorber mechanism}} has arisen from
deeper considerations which were not themselves concerned with mechanisms at
all.

\subsection{Overview of Results}
The theory, as we have so far developed
it, is quasi-classical insofar as it is a one-clock model. This restriction is
not structural - as we point out in Roscoe \cite{RoscoeC} - but was imposed in
order not to obscure the central arguments of the original development. However,
it does mean that any gravitation theory which is based upon this particular
formal development of inertial space \& time can only be applicable to
weak-field regimes. Hence, here we restrict ourselves to showing how classical
Newtonian theory is reproduced, and to the application of the theory to model an
idealized spiral disc - one without a central bulge and with perfect cylindrical
symmetry in the disc. A particular subclass of solutions, corresponding to
logarithmic spirals, predicts that the circular velocities should behave
according to $V = AR^{\,\alpha}$ where $(A, \alpha)$ are parameters which vary
between discs, and where $\alpha$ must necessarily satisfy a certain algebraic
consistency condition.  In the first instance,
we ignored this consistency condition and focussed on the basic power-law model.
This simplified model has been extensively tested on several very large samples
(900, 1182, 497 and 305 objects respectively) in Roscoe \cite{RoscoeA} and
Roscoe \cite{RoscoeD}, and shown to resolve the data with a remarkable
statistical precision.

However, in the course of the initial study, a major new phenomenology was
discovered - that of {\it discrete dynamical classes} in spiral discs.
This has been the subject of an initial study, Roscoe \cite{RoscoeB},
and a comprehensive later study involving the four large samples,
Roscoe \cite{RoscoeD}, which confirmed the phenomenology at the level
of statistical certainty.

This latter discovery caused us to reconsider the initially ignored
$\alpha$-consistency condition, and we found that its algebraic form provides
all the structure required to understand the {\it discrete dynamical classes}
phenomenology. Thus, we can potentially understand the new phenomenology as the
physical manifestation of an algebraic consistency condition that must
necessarily be satisfied in order for solutions of the complete system to exist.

Finally, we apply the theory with complete success to the detailed modelling of
a sample of eight Low Surface Brightness spirals (LSBs). The point of choosing
such objects is primarily because they are so diffuse that, according to the
canonical viewpoint, they must consist of $>99\%$ dark matter to be
gravitationally bound. This is to be compared with, typically,  $>95\%$ dark
matter for ordinary spirals.

\subsection{The MOND Connection} Finally, we infer
from the succcess of the theory that it must intrinsically provide a theoretical
basis for the MOND (Modified Newtonian Dynamics) programe of Milgrom
\cite{Milgrom1983a}, \cite{Milgrom1983b}, \cite{Milgrom1983c}. In considering
the flat rotation curve problem of spiral galaxies, Milgrom had the idea that,
perhaps, in extremely weak gravitational fields ($g<<10^{-10}ms^{-2}$), the
nature of Newtonian gravitational mechanics changed in such a way that flat
rotation curves were the natural result. The notable thing about MOND is that,
whilst it was designed to address one particular phenomenology - that of the
flat rotation curve in galaxy discs - it has enjoyed impressive success in a
variety of quite distinct circumstances, making several preditions that have
subsequently been verified - and any one of which could have falsified the
theory. On any objective measure, the performance of the one-parameter MOND
model is superior to that of the multi-parameter Cold Dark Matter model. See for
example, de Blok \& McGaugh \cite{deBlok1998} and McGaugh \& de Blok
\cite{McGaugh1998a}, \cite{McGaugh1998b}. The primary difficulty for the MOND
programme has been that there is no underlying theory to support it. By virtue
of the present theory's complete success in modelling the LSB sample, we infer
that it must contain the quantitative essence of the MOND algorithm.
\section{Review of Fundamental Arguments} Although the mathematical machinery
used in  Roscoe \cite{RoscoeC} is essentially straightforward, the ideas and
forms of argument used in that development will be unfamiliar to most readers.
Therefore, a short review of the process employed will probably be useful here.

The basic aim was to give expression to a particular form of Mach's Principle -
essentially the idea that within the framework of a properly fundamental theory
it should be impossible to conceive {\it empty} physical space \& time. This was
achieved, briefly, via the following process:
\begin{itemize}
\item  Note that, on a large enough scale ($ > 10^8$ lightyears, say) it is
possible, in principle, to write down an approximate functional relation giving
the amount of mass (determined by {\it counting} particles) contained within a
given spherical volume, $M\approx F(R)$; \item  Since $M$ will be a monotonic
function of $R$, we can invert this latter expression to get $R\approx G(M)$;
\item Whilst it is conventional to suppose that $M \approx F(R)$ is logically
prior to $R\approx G(M)$, there is no natural imperative dictating that we {\it
cannot} reverse the logical priorities;
\item That is, we are at liberty to suppose that $R\approx G(M)$ is the prior
relationship so that, in effect, the radius of a sphere is defined
in terms of the amount of mass contained within it. In other words, we have made
it impossible to define a radial measure in the absence of material. This is the
first critical step towards the Machian theory required;
 \item Once we have a
radial measure, defined in terms of $M$, we can define a coordinate system
allowing us to label points in the space and to specify displacements in the
space. But we have no means of associating an {\it invariant length} to any such
displacement - there is no metric yet;
\item To obtain a qualitative idea of metric, we made the
following thought experiment: An observer floats without effort through a
featureless landscape. But this observer will have no sense of distance
travelled - there is no metric. By contrast, suppose now that the  landscape
possesses many distinctive landmarks.
Now there will be a powerful sense of distance travelled imposed by the
continually changing relationships between the landmarks and the observer. In
other words, the observer's {\bf \emph{changing perspective}} of the landscape
provides him with the means to make qualitative judgements of  the magnitudes of
his displacements in that landscape. We are able to use this idea to obtain a
quantitative definition of a metric within the model universe described below;
\item Define a model universe consisting of particles ({\it not} assumed to be
in a static relationship to each other) possessing {\it only} the property of
enumerability (we have in mind that {\it mass} is fundamentally a measure of the
{\it amount} of material, and therefore determinable by a counting process), and
suppose that there is at least one point about which the distribution is
spherical; \item There is no notion of {\it time} yet but, even so, we have to
distinguish between the possibilities of {\it non-evolving} and {\it evolving}
model universes. This distinction turns out to be the distinction between a
non-relativistic universe and a relativistic one and, to simplify matters in the
first instance,  we supposed that the model universe was non-evolving; \item The
latter two points give $R=G(M) \rightarrow M=F(R)$. The particles within any
given level surface of $M$ are then taken to define the landmarks within our
landscape, and a straightforward modelling process then allows us to use the
{\bf \emph{changing perspectives}} idea above to define a metric within the
model universe. We found: \begin{displaymath} g_{ab} \equiv \nabla_a \nabla_b
{\cal M} \equiv {\partial^2 {\cal M} \over \partial x^a \partial x^b} -
\Gamma^k_{ab} {\partial {\cal M} \over \partial x^k},
\end{displaymath} where ${\cal M}$ is a simple linear function of $M(R)$, and
$\Gamma^k_{ab}$ is chosen to be the metric affinity. This choice is made because
it guarantees that appropriate generalizations of the divergence theorems exist
- which is necessary if we are to have conservation laws in the model universe.
\end{itemize}
\section{Gravitation: General Comments}
\subsection{Newtonian Theory and Point-Mass Perturbations}
We have stated, in \S\ref{Sec1}, that ${\cal M}$ in the $D=2$ equilibrium
universe can be properly considered as a classical representation of a material
vacuum within which the particles (or quasi-photons) arbitrate between length
and time scales in a way which is reminiscent of Bondi's interpretation of $c$.

But what about gravitational processes in such a universe? If they exist at
all, they can only arise via  perturbation processes in the material vacuum
generated by conventionally understood point-mass sources. Given this, and
thinking crudely in terms of the quasi-photons each moving with a speed $v_0$
(remember, $v_0$ is actually a conversion factor having dimensions of speed), it
quickly becomes apparent that {\it gravitational effects} between two such
point-masses are most readily understood in terms of each point-mass partially
shading the other from vacuum particle collisions, so that each picks up a
net momentum towards the other, as if attracted by a gravitational force. Of
course, the emergence of Newtonian Gravitation for the case of the single
point-mass perturbation is the first necessary condition that must be met, if
the foregoing picture is to be given credence. This condition is shown to be met
in appendices \S\ref{Prelim}, \S\ref{Sec2} and \S\ref{Sec3}.

The $N$-body  theory is developed by considering the perturbations generated in
the material vacuum by finite ensembles of point-mass particles in appendices
\S\ref{GMC}, \S\ref{TBP}, \S\ref{NBP} and \S\ref{Sec9}. However, in the course
of this development, a subtlety arises concerning the interpretation of ${\cal
M}$: specifically, in the equilibrium universe, ${\cal M}(r)$ simply describes
the distribution of vacuum mass about any centre (it is fractal, $D=2$). But,
when the equilibrium universe becomes perturbed by a finite ensemble of
point-mass particles, then:
\begin{itemize}
\item In order to describe the changing states of motion of a
specified point-mass particle, the vacuum mass {\lq seen'}
by that particle (that is, that component of vacuum mass which acts to cause a
change in motion)  must be defined;
\item
The unique association of a specified point-mass particle with a particular
vacuum mass distribution is accomplished in the analysis of \S\ref{GMC}, which
shows how the requirement for linear momentum conservation within the finite
point-mass ensemble implies that the vacuum mass {\lq seen'} by a particle of
mass $m$ at position ${\bf r}$ (defined with respect to the ensemble mass
centre) must have the functional form ${\cal M} \equiv {\cal M}(m{\bf r})$.
Equivalently, this implies that the vacuum mass {\lq seen'} by a point-mass
particle of {\it unit} mass at position ${\bf r}$ is given by ${\cal M}({\bf
r})$. \end{itemize}
\subsection{Continuum Perturbations: A Necessary Conjecture}
\label{Conjecture}
We show, in appendix \S\ref{Sec2}, how a perfectly workable quasi-Newtonian 
point-source theory arises from a point-mass perturbation of the equilibrium
vacuum. However, our primary interest is in gravitational processes within
extended material distributions. But, for these, we have no means of giving a
quantitative definition to  the corresponding vacuum mass distribution,  ${\cal
M}({\bf r})$ - we have only been able to write down a probable general structure
in \S\ref{NBP}. Therefore, we must necessarily rely upon the
following qualitative arguments:

Bearing in mind that the vacuum mass {\lq seen'} by a
particular point-mass particle is actually some measure of the {\it discrepancy}
between the $D=2$ equilibrium distribution and the perturbed distribution, and
that this discrepancy is generated by the finite point-mass ensemble, then the
vacuum mass {\lq seen'} by a particular point-mass particle must also be a
measure of the total of other point-masses within the ensemble.
This leads us to the following conjecture:
\begin{quote}
In the limit of the finite
point-mass ensemble becoming a continuum distribution, then  this distribution
traces the vacuum mass distribution {\lq seen'} by a test particle to the extent
that one can act as a proxy for the other. \end{quote}
As we shall see in the later sections, this
conjecture (or something very similar) is strongly supported  on the data for
Low Surface Brightness (LSB) galaxies analyzed in later sections.
\section{A Model For An Idealized Spiral Galaxy} In the following,
we consider the application of this theory to model  an idealized spiral galaxy -
defined to be one without a central bulge, and without the irregularities that
are routinely present in the discs of real galaxies, and without the bars that
are present in many spirals. We model this idealized spiral
galaxy as a mass distribution possessing perfect cylindrical symmetry.

In order to motivate the detailed analysis of \S\ref{ISG1} and \S\ref{ISG2} and
the appendices more strongly, we begin by considering the solutions which arise
and the high quality of their fit to the data. Furthermore, because, in our
view, the work presents a very interesting example of the idealized cycle {\it
theory $\rightarrow$ test $\rightarrow$ discovery $\rightarrow$ more theory
$\rightarrow$ more test},  we present our solutions in chronological order of
their calculation to illustrate the way in which the work as a whole has
progressed. \subsection{The Power-Law Model}
\label{SLS}
In the first instance, we were only able to progress by making  use of the
empirical knowledge that the spiral structure of spiral galaxies is essentially
logarithmic. This  led directly to the idealized disc solutions:
\begin{equation} V_{rot} =
AR^\alpha,~~~~V_{rad} = BR^\alpha,~~~~\rho=\frac{C}{R^2} \label{LinSoln}
\end{equation} where $R$ is the radial position within any given galaxy disc,
$V_{rot}$ is the rotational velocity,  $V_{rad}$ is the radial velocity,
$\rho$ is the mass density, $A,B$ and $C$ are arbitrary constants, and $\alpha$
is constrained to satisfy \begin{equation}
\alpha=1-\frac{p}{Sq}-\frac{p^2+q^2}{qK_1} ~~{\rm where}~~
p\,K_1^2\,\pm\,(1\,+\,p^2\,+\,q^2)\,K_1\,+\,p\,=\,0,
\label{AlphaConstraint} \end{equation} where $(p,q)$ are undetermined constants
and $S=\pm1$.
This solution (ignoring the constraint equations for $\alpha$)  led directly
to the data analysis described in detail in Roscoe \cite{RoscoeA} and described
briefly in \S\ref{SecPowerLaw}.
\subsection{The $\alpha$-Constraint}
\label{ParticularSoln}
However, the data analysis of Roscoe \cite{RoscoeA} led, in its turn, directly
to the discovery of the {\it discrete dynamical states} phenomenology in galaxy
discs, briefly reviewed in \S\ref{DDC}. This phenomenology was  reported
initially in Roscoe \cite{RoscoeB} for the case of one large data set, and
subsequently confirmed in Roscoe \cite{RoscoeD} for three further large data
sets. This discovery led us to reconsider the role of the $\alpha$-constraint
(\ref{AlphaConstraint})
which we had previously ignored.

\begin{figure} \begin{center}
\resizebox{10cm}{!}{\includegraphics{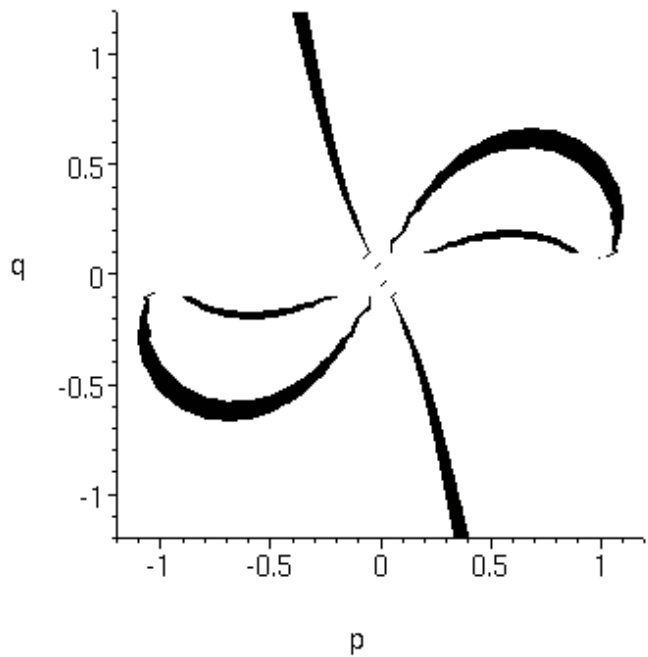}} \caption{\label{figH}
Intersection of three surfaces in $\alpha=0.5$} \end{center} \end{figure}

Bearing in mind that there are four possible solutions for $K_1$ in terms of
$(p,q)$ and that the expression for $\alpha$ contains $S=\pm 1$ then, at face
value, there are eight possibilities fo $\alpha$. However, a closer
investigation shows that there are only six {\it distinct} possibilities, so
that we have the general structure
\begin{equation} \alpha =
F_i(p,q),~~i=1\, .. \,6 \label{AlphaCond} \end{equation} Thus, in principle, any
given galaxy disc is potentially associated with one of six distinct surfaces in
$(\alpha,p,q)$ space.   Figure \ref{figH} shows the intersection of three of
these surfaces with the plane $\alpha=0.5$. The three omitted surfaces are
mirror images (about either axis) of those shown so that the
total figure is perfectly symmetric about the two axes. (The poor figure quality
is a function of the limitations of the Maple graphics package).
We show in \S\ref{DDC} that the surfaces (\ref{AlphaCond}) provide all the
structure required to arrive at a qualitative understanding on the discrete
dynamical states phenomenology.
\subsection{General Solutions} \label{GSCS}
The big open question concerning spiral galaxies is whether dark matter exists
or not and, amongst the class of spiral galaxies, it is the Low
Surface Brightness (LSB) galaxies which present the most extreme problems.
Typically, these objects are estimated to consist of more than $99\%$ dark
matter, according to the canonical viewpoint.
Therefore, it is these objects which must be successfully modelled by any new
theory if that theory is to have maximum credibility.

Numerical techniques appear to offer the only realistic possibility for
modelling specific galaxies. But, there are serious problems even here:
specifically, the first galaxy in the LSB class was only discovered in the late
1980s and, even today, there are only a few tens of them with detailed
and accurate rotation velocity measurements. Of these, only a minor fraction
have detailed estimates of the ordinary matter in their discs and none
have any radial flow measurements. In general, we would therefore expect that
the detailed modelling of any specific LSB would necessarily require sweeping
assumptions to be made about radial flows - thereby considerably reducing the
value of any such modelling process (especially so when the equations concerned
are numerically stiff).

However, in the case of the present theory, it transpires that the radial
velocity component can be algebraically {\it eliminated} from the complete
system, leaving a reduced system in rotation velocity and mass density only.
\subsubsection{Mass Distribution: The Beautiful Equation}
\label{MDBE}
Not only is it possible to eliminate the radial velocity explicitly from the
complete system, but the mass equation can be expressed in a form which is
independent of velocity at all! In this form, it is given by:
\begin{equation}
(\psi^2-1)\epsilon^2+\frac{2}{q}\left[p(1+\psi^2)+(1+p^2+q^2)S\,\psi\right ]
\epsilon -(\psi^2-1)=0\,, \label{Mass1} \end{equation}
\begin{equation}
{\rm where}~~~\epsilon^2 \equiv -1 -
\frac{R}{\rho}\,\frac{d\rho}{dR}~~~{\rm and}~~~\psi^2 \equiv 1-\frac{2 \pi
\rho}{m_0} \,R^2.
\label{Mass2}
\end{equation}
Here, $S=\pm1$ whilst the parameters $(p,q,m_0)$ are integration constants. Of
these $(p,q)$ are as in \S\ref{SLS}, whilst $m_0$ can be fixed to
have a magnitude of $4 M_{gal}$ where $M_{gal}$ is the mass of the object being
modelled estimated from measurements of stars, gas and dust. Since this is
fixed independently of any calculations then there are, in effect, only two
disposable parameters, $(p,q)$, for the mass and circular velocity equations. It
is the algebraic structure of the mass equation above which, ultimately,
allows the theory to model the various complexities manifested by rotation
curves as a class, and which leads us to refer to it as {\it the beautiful
equation}.

At face value - and when the signature $S=\pm1$ is taken into account - equation
(\ref{Mass1}) is satisfied by any one of four distinct distributions for $\rho$.
However, the situation is more complex than this: reference to (\ref{Mass1})
shows that if $\epsilon = \pm1$, then $\psi$ assumes one of two possible
constant values. It then follows, by the second of (\ref{Mass2}), that $\rho
\sim 1/R^2$. But any $\rho$ of this form also satisfies the first of
(\ref{Mass2}). The net result is that the mass equation, (\ref{Mass1}), is
satisfied when $\rho$ satisfies any one of {\it six} possible distributions.

However, closer analysis then reveals that of the two possibilities associated
with $\epsilon =\pm1$, one is not physical (it corresponds to negative
densities) and of the four possibilities associated with $\epsilon \neq \pm1$,
a further one is also not physical. The net effect is that
(\ref{Mass1}) is satisfied by any one of four physically realizable
distributions for $\rho$  - so that, in principle, the distribution of material
in any given galaxy disc can satisfy different differential equations over
different anular sections. For example, the single admissable $\rho$
distribution corresponding to $\epsilon=\pm1$ actually corresponds to the
solution of \S\ref{SLS}, and we find that six of our
sample of eight LSBs have such segments embedded within their discs.
\subsubsection{The Rotation Velocity Distribution} The rotation velocity
equation is given by given by \begin{equation} \frac{1}{V}\frac{dV}{dR} =
\frac{1}{R}\left[\frac{1}{2}(1+\epsilon^2)-\frac{p}{q}\,\epsilon -
\left(\frac{p^2+q^2}{q}\right)\,\frac{\epsilon}{S\,\psi}\right] \label{VelEqn}
\end{equation} where we note that $\rho$ appears through $\epsilon$ and $\psi$.

In practice, we have found that one (or two)  switches between different
$\rho$ distribution laws - and hence between different velocity distribution
laws - occur somewhere in every LSB disc, and it is this mechanism which allows
the theory to model the complex behaviour of rotation curves. We show the
results of a detailed modelling exercise on a sample of eight LSBs in
\S\ref{LSBs}.

\subsubsection{General Comments}
It is interesting to note that, if $V=V^*$ is a solution of (\ref{VelEqn}), then
so is $V=kV^*$, for any constant $k$. This is a useful property in the present
context since, because galaxy discs are generally not seen edge on, we can only
estimate rotation velocities when we have estimated disc-inclination - and the
effect of getting this wrong is simply to scale the true rotation velocities by
an  unknown constant factor. Thus, (\ref{VelEqn}) is indifferent to knowledge
about disc-inclination angles.

Finally, we note that the existence of a switching mechanism in the theory is
reminiscent of the similar thing which is a necessary component of the
MOND algorithm, discussed in later sections.
\section{Power Law Dynamics: The Observations}
\label{SecPowerLaw}
\subsection{General Comments}
In this
section, we give a brief review of the published evidence supporting the view
that velocity distributions in {\it idealized} logarithmic discs behave
according to the basic power law solutions, (\ref{LinSoln}). There are four
preliminary comments: \begin{itemize}
\item Firstly, it is essential to understand that, so far as the phenomenology
is concerned, we are only talking about rotation curves over their interior
segments on which they are rising strongly - that is, we are explicity excluding
the exterior flat parts. As it happens, practical considerations ensure that
{\it optical} rotation curves are generally confined to this region anyway, and
the analyses which follow are all confined to optical rotation curves (ORCs).
\item Secondly, whilst a large amount of $V_\theta$ (circular
velocity) data exists in the form of several large samples of published ORCs,
there is no corresponding body of data for radial velocity flows. The reason is
simply that such flows are, typically, an order of magnitude smaller than the
circular flows and the techniques to measure them have only recently become
available. \item Thirdly, our analysis applies only to idealized discs defined
to possess perfect cylindrical symmetry. Since spiral galaxies typically possess
bulgy central regions then our model can, at best, only have validity in those
parts of the disc which are {\it exterior} to the innermost central regions and
(by the initial comment) {\it interior} to the very outermost regions where the
rotations curves become flat.  \item Fourthly,
since galactic discs are generally complete with all manner of irregularities
then the model can only have a {\it statistical} validity. It is for this reason
that we confine our selves to the analysis of very large samples only.
\end{itemize}
\subsection{The ORC Samples}
Mathewson, Ford \& Buchhorn \cite{Mathewson1992} published a
sample of 900 optical rotation curves (ORCs) which provided the basis for our
first large scale analysis of disc dynamics from a power-law point of view. The
results of this analysis, (Roscoe \cite{RoscoeA}) showed that the power-law
resolves disc dynamics in the outer part of optical discs to a very high degree
of statistical precision. We have subsequently analysed three further large
samples, these being those of the 1182 ORCs published by Mathewson \& Ford
\cite{Mathewson1996}, the 497 ORCs published by Dale et al (\cite{Dale1997} et
seq) and the 305 ORCs published by Courteau \cite{Courteau}. This last sample
differs from the previous three in being the only one using $R$-band photometry
rather than $I$-band photometry. For associated technical reasons, the modelling
process for the Courteau sample differs in its details from the others and so is
excluded from the first part of the present discussion.
\subsection{The Basic $(\alpha,\ln A)$ Plot}
\begin{figure}
\noindent
\begin{minipage}[b]{\linewidth}
\resizebox{12cm}{!}{\includegraphics{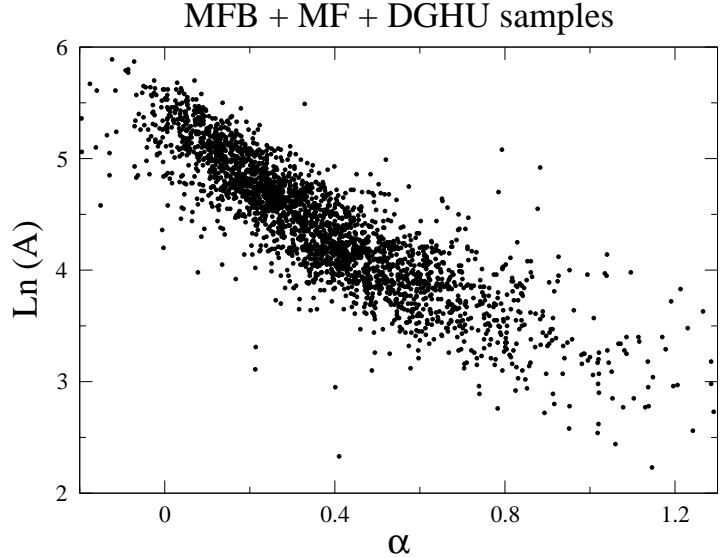}}
\end{minipage}
\caption{Plot of $(\ln A,\,\alpha)$ for 2405 galaxies}
\label{fig1}
\end{figure}

The basic question is: does the power-law $V_{rot} = A R^{\,\alpha}$ provide
a good resolution of ORC data on the exterior part of optical discs\,?
(Remember, practical considerations ensure that ORCs generally to not extend
to the far exterior regions where rotation curves become flat). The first
problem here is to give an objective definition of what is meant by the {\it
exterior part of the optical disc}. This is provided originally in Roscoe
\cite{RoscoeA}, and more clearly in Roscoe \cite{RoscoeD}. Once this is done,
the analysis uses linear regression to estimate the parameter pair $(\ln
A,\,\alpha)$ for each of the ORCs in the sample. Figure \ref{fig1} gives the
scatter plot for all of the 2405 usable ORCs in the three $I$-band samples.
(About 7\% of the total sample of 2579 ORCs were lost to the analysis because of
objective reasons associated with data quality.) 
There are three significant points to be made:
\begin{itemize}
\item The first point, which came as a shock, is the existence of a very
clear and powerful correlation between $\ln A$ and $\alpha$, since there is no
obvious apriori reason why any correlation should exist at all;
\item Secondly, although there is no apriori reason to expect an $(\alpha,\ln
A)$ correlation, equation (\ref{AlphaConstraint}) states that $\alpha$ is
correlated with the dynamical parameters $(p,q)$. But, by definition, $\ln
A$ is strongly determined by the dynamics and so it follows that, in
qualitative terms, the $(\alpha,\ln A)$ correlation of figure \ref{fig1} is
consistent with the existence of a relation like the
$\alpha$-constraint of (\ref{AlphaConstraint}); \item Thirdly, except for
what is probably statistical scatter, $\alpha$ appears to be strongly confined
in the region $\alpha \in(0,1)$. In section \S\ref{GCSTT}, we see how this
$\alpha$ confinement has an elegant association with certain topological
transitions in the $(\alpha,p,q)$ parameter space. \end{itemize}

\subsection{A Detailed Model of the $(\alpha,\ln A)$ Plot}
\label{DMALP}
A more detailed analysis of this diagram showed that the luminosity properties
of galaxies vary very strongly through the plot. Specifically, consider the
power-law model, $V = A\,R^{\,\alpha}$, in dimensionless form:
\begin{equation}
{V \over V_0}\,=\,\left({R \over R_0}\right)^\alpha~~\rightarrow~~
A\,=\,{V_0 \over R_0^{\,\alpha}}\,. \label{eqn6}
\end{equation}
Then,
defining $M$ as the absolute $I$-band magnitude and $S$ as the
absolute $I$-band surface brightness of an object, a detailed modelling of
the MFB (Mathewson et al \cite{Mathewson1992}) and MF
(Mathewson \& Ford \cite{Mathewson1996}) data in
figure \ref{fig1} shows that the particular model
\begin{eqnarray}
\ln A &=& \ln V_0 - \alpha \, \ln R_0, \nonumber \\
\ln  V_0 &=& -1.596 - 0.316 \,M  \label{7B} \\
\ln R_0 &=& -7.614\, -0.474\,M -0.0050\,S 
\nonumber
\end{eqnarray}
accounts for about 93\% of the total variation in the figure.
It is to be noted that, in the model, the $t$-statistic for each of the model
parameters
satisfies $|t| > 11$, so that all the included variations are powerfully present.
This is, in itself, a strong demonstration of how effectively the power-law
model resolves ORC data. 
The model fit can be improved by including the DGHU (Dale et al \cite{Dale1997} et seq)
data in the modelling process -
but we choose to exclude it to provide an independent test, discussed below.
\subsection{An Alternative Visualization of the  Model-Fit}
A very effective alternative way of visualizing the fit of the model
(\ref{7B}) to the data can be obtained as follows: Suppose we use the
definitions of (\ref{7B}) in the dimensionless form give at (\ref{eqn6}),
so that the measured $(R,\,V)$ data for each ORC is scaled by the luminosity models
for $(R_0,\,V_0)$ for that ORC, and then regress
$\ln (V / V_0)$ on $\ln ( R / R_0)$ for each ORC.
Then, if the power-law model (\ref{eqn6}) is good, we should find a null zero 
point for each ORC - except for statistical scatter.

\begin{center}
\begin{figure}[h]
\noindent
\resizebox{15cm}{!}{\includegraphics{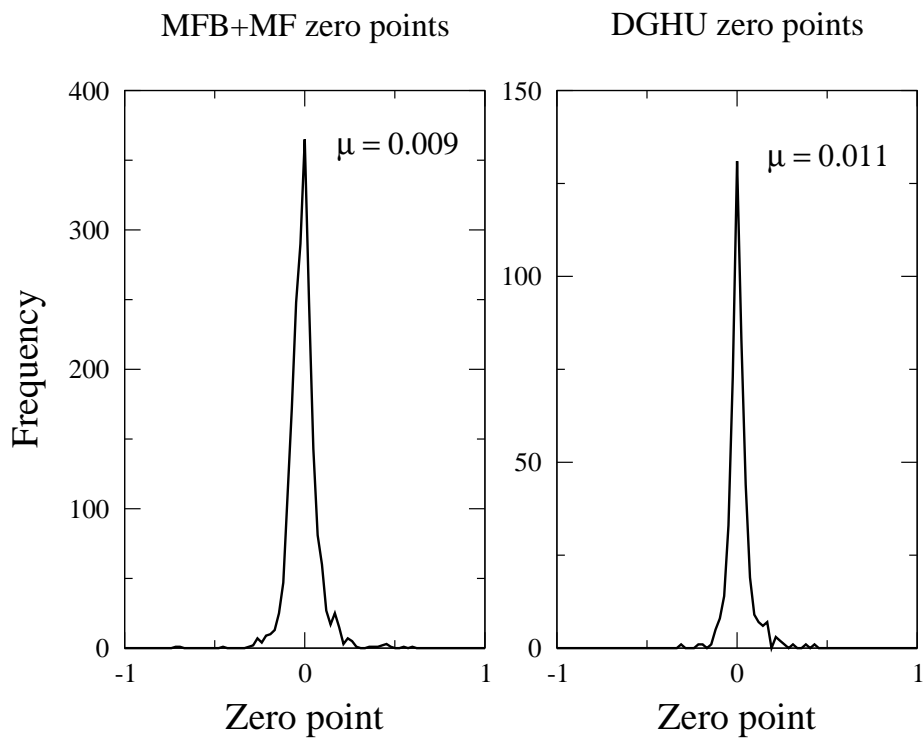}}
\caption{Plot of zero point for 1951 MFB+MF galaxies and 454 DGHU galaxies}
\label{fig15}
\end{figure}
\end{center}

Figure \ref{fig15} (left) gives the frequency diagram for the actual zero
points computed for the combined Mathewson et al (\cite{Mathewson1992},
\cite{Mathewson1996}) samples from
which the model (\ref{7B}) was derived, whilst a wholly
independent test of the model is given by figure \ref{fig15} (right) which
gives the 
frequency diagram for the zero points derived from the DGHU sample using the
model (\ref{7B}) - which, of course, was derived {\it without} DGHU data.
It is clear that there is absolutely no evidence to support
the idea that these zero points are different from the null
position. Thus, the power-law model is strongly supported, and
(\ref{eqn6}) with (\ref{7B}) can be considered to give a high precision
statistical resolution of ORC data in the exterior part of the disc.
\begin{figure}[p]
\resizebox{14cm}{!}{\includegraphics{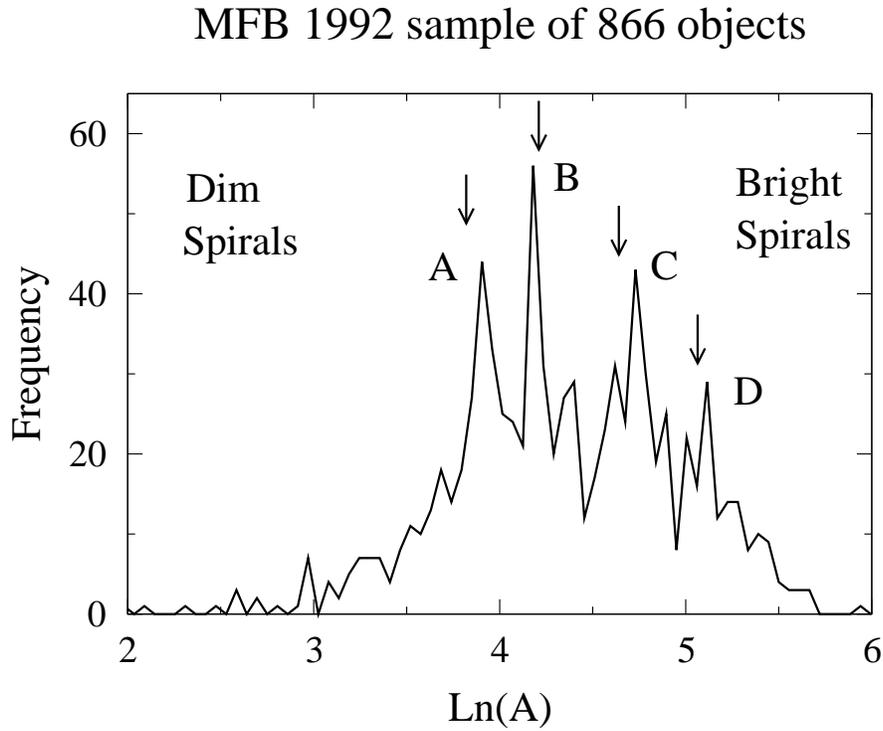}}
\caption{Vertical arrows represent predicted positions of peaks from an
analysis of 12 Rubin et al ORCs}
\label{fig3}
\end{figure}
\begin{figure}[p]
\resizebox{14cm}{!}{\includegraphics{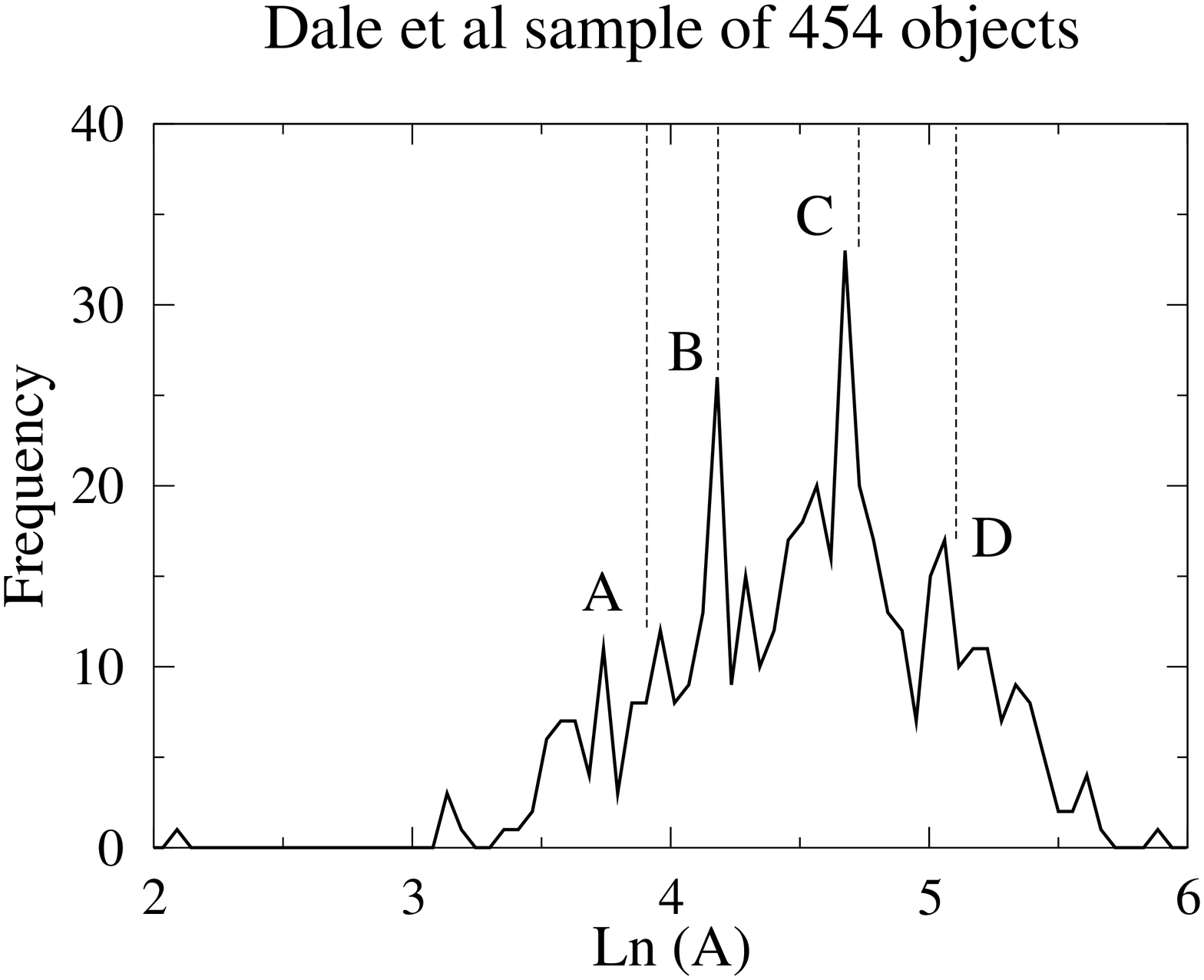}}
\caption{Vertical dotted lines indicate peak centres of
figure \ref{fig3}.}
\label{fig4}
\end{figure}
\begin{figure}[p]
\resizebox{13cm}{!}{\includegraphics{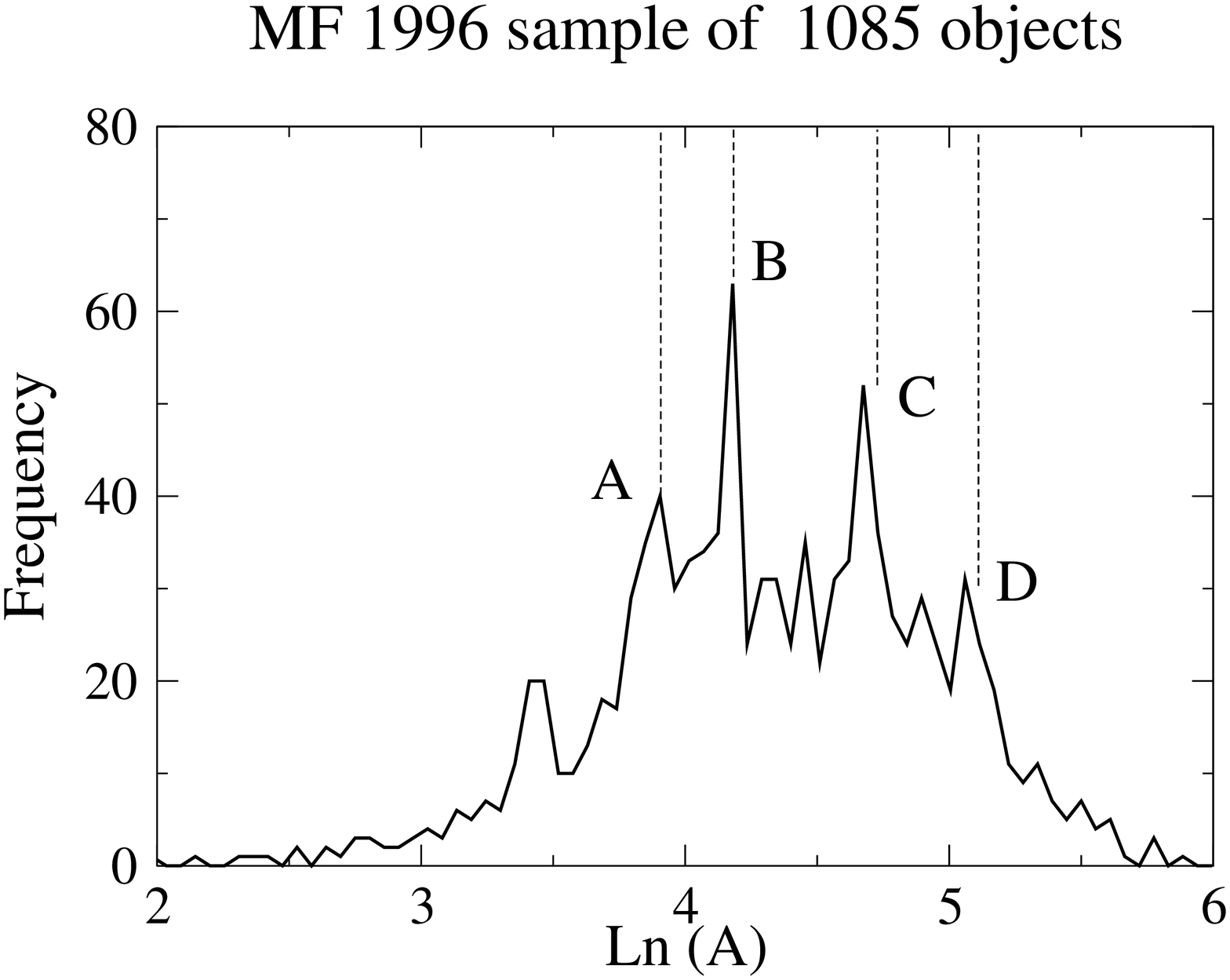}}
\caption{Vertical dotted lines indicate peak centres of
figure \ref{fig3}.}
\label{fig5}
\end{figure}
\begin{figure}[p]
\resizebox{13cm}{!}{\includegraphics{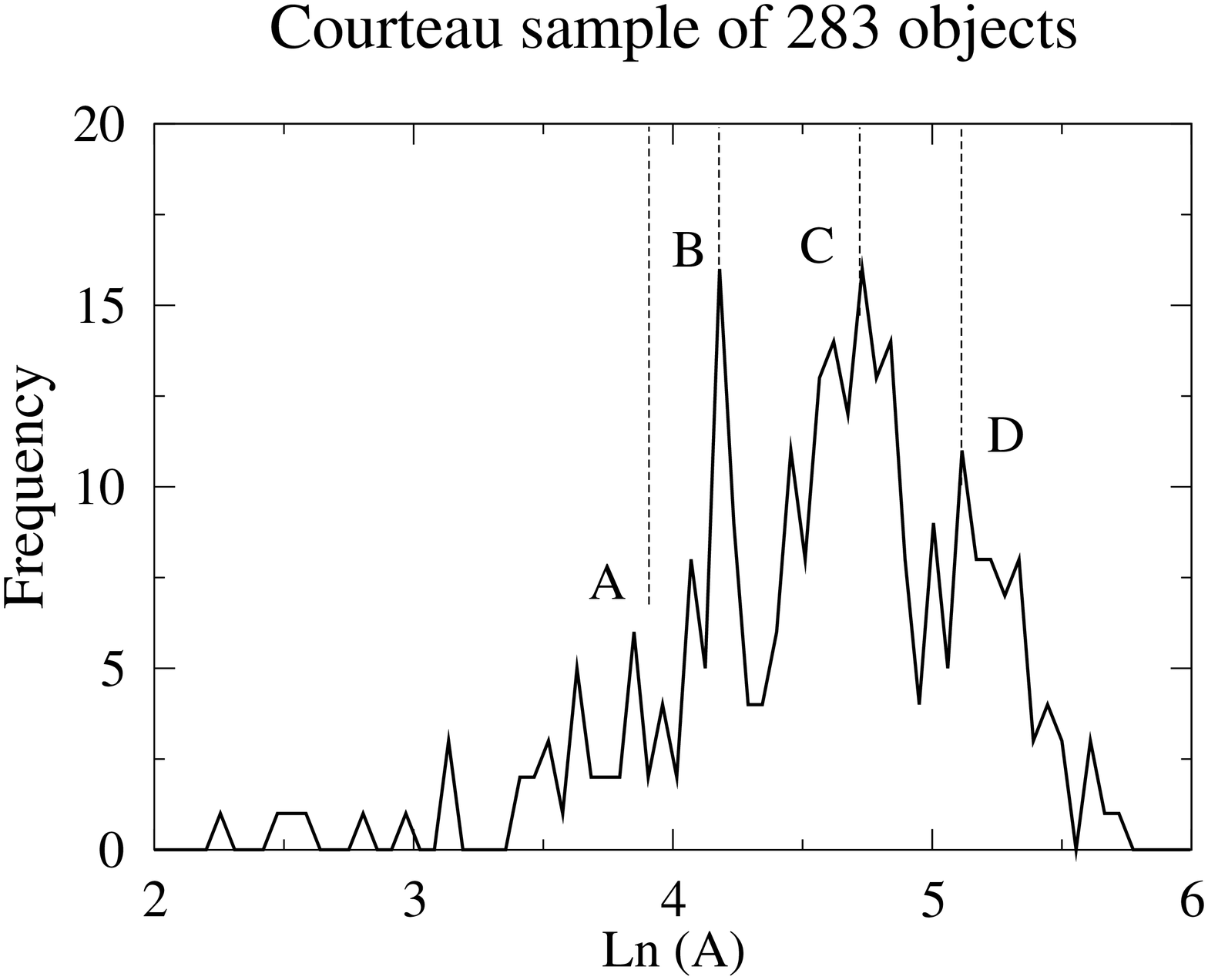}}
\caption{Vertical dotted lines indicate peak centres of
figure \ref{fig3}.}
\label{fig6}
\end{figure}

To summarize, we have so far shown that the rotational part of the power-law
model (\ref{LinSoln}) is very strongly supported on the data and have indicated 
that the data does not yet exist to consider the validity - or otherwise - of
the radial component of the model.
\section{Discrete Dynamical Classes: The Observations}
\label{DDC}
\subsection{General Comments}
In this section we consider the exact solutions of
\S\ref{SLS} - specifically, we consider the extremely strong
statement, made at (\ref{AlphaCond}), to the effect that $\alpha$,
calculated for any given spiral, is constrained to occupy one of a set of
distinct surfaces in $(\alpha,\,p,\,q)$ space, where $(p,\,q)$ are dynamical
parameters in the model.

This statement has interest on many levels - not least because
nothing like it features in any other extant theory of disc dynamics.
The effect was first noticed - tentatively and prior to us recognizing the
potential significance of the $\alpha$-constraints - during a pilot study of a
small sample of Rubin et al \cite{Rubin} ORCs, and this initial identification
was used to define a hypothesis which was subsequently tested on the Mathewson
et al \cite{Mathewson1992} sample, and reported in Roscoe \cite{RoscoeB}. The
effect was subsequently confirmed in Roscoe \cite{RoscoeD} on three further
large samples (Dale et al \cite{Dale1997} et seq, Mathewson \& Ford
\cite{Mathewson1996} and Courteau \cite{Courteau}). For completeness, we give a
brief review of this evidence here.

In practice, the evidence takes the form of the $\ln A$ frequency diagrams for
each of the four samples, and then interpreting the meaning of these
diagrams. The computation of the $\ln A$ parameter is as it
was for figure \ref{fig1} (but see Roscoe \cite{RoscoeD} for a complete discussion).
\subsection{The $\ln A$ Frequency Diagrams}
Figure \ref{fig3} shows the $\ln A$ distribution arising from the analysis
of the Mathewson et al \cite{Mathewson1992} sample, and the short vertical arrows in that
figure indicate the predicted positions of the peaks, based on a pilot study
of a sample of twelve ORCs from Rubin et al \cite{Rubin}. Given the very small size
of this initial sample of Rubin objects, we can see that the match is remarkable.

Figures \ref{fig4}, \ref{fig5} and \ref{fig6} show the corresponding
distributions for the Dale et al (\cite{Dale1997} et seq) sample,
the Mathewson \& Ford \cite{Mathewson1996} sample and the Courteau
\cite{Courteau} sample. In each of these cases, the vertical dotted
lines indicate the peak centres of figure \ref{fig3}.

The $A$ peak in figures \ref{fig4} and \ref{fig6} are more-or-less absent
because this peak corresponds to very dim objects, and these are very much
under-represented in the two samples concerned.

The joint probability of the observed peaks in the four samples arising by
chance alone, given the
original hypothesis raised on the small Rubin et al sample, has been computed
in Roscoe \cite{RoscoeD}, using extensive Monte-Carlo simulations, to be vanishingly
small at $\approx 10^{-20}$.
\subsection{Interpretation of the $\ln A$ Frequency Diagrams}
It is clear from the four diagrams that $\ln A$ has a marked preference for
one of four distinct values, say $\ln A \,=\, k_1,\,k_2,\,k_3,\,k_4$.
However, we also know, from \S\ref{DMALP}, that $\ln A$ is a
strongly defined function of the galaxy parameters $(\alpha,\,M,\,S)$, so that
$\ln A = F(\alpha,\,M,\,S)$.
Putting these two results together gives
\begin{equation}
F(\alpha,\,M,\,S) \,=\, k_i\,,~~~~i = 1,2,3,4\,.
\label{eqn12}
\end{equation}
Consequently, the $\ln A$ frequency diagrams imply that spiral discs are
confined to one of four distinct surfaces in $(\alpha,\,M,\,S)$ space.

But $M$ is a measure of absolute galaxy luminosity, and therefore a measure
of the corresponding total galaxy mass, $m$ say (assuming no dark matter).
Similarly, the surface brightness parameter, $S$, which is a measure of the
density of absolute galaxy luminosity, can be considered as a measure of
mass density, $\rho$, in a galaxy. That is, $(M,\,S) \approx (m,\,\rho)$ and
these latter two parameters are, apriori, important dynamical parameters for
any given system. If we now presuppose the existence of a mapping
$(m,\rho)\rightarrow (p,q)$ where $(p,q)$ are the dynamical parameters in our
model equations, then we see that (\ref{eqn12}) is consistent with
(\ref{AlphaCond}) - the difference being that (\ref{AlphaCond}) allows up to
six surfaces, whereas we have only identified four in the phenomenology.

We reasonably state, therefore, that the theory contains all the structure
required to explain, {\it in principle}, the phenomenon of discrete dynamical
classes for disc galaxies - subject to the existence of appropriate mappings
$(m,\rho)\rightarrow (p,q)$. However, various significant details uncovered in 
the analysis of \S\ref{LSBs} give us further insight into the problem, and these
are briefly discussed in \S\ref{MODDCP}.
\section{Detailed Modelling of LSB Galaxies}
\label{LSBs} In this section, we apply equations (\ref{Mass1}) and
(\ref{VelEqn}) of \S\ref{GSCS} to the detailed modelling of a sample of eight
LSB galaxies \footnote{Provided by Stacy McGaugh, of the University of Maryland,
USA} which have already been successfully modelled by MOND (deBlok \& McGaugh
\cite{deBlok1998} and McGaugh \& deBlok \cite{McGaugh1998b}). We shall not dwell
on the computational problems involved in solving these equations, but we shall
discuss, briefly, the means by which the parameters $(m_0,p,q)$ were determined.
\begin{figure} \begin{center}
\resizebox{12cm}{!}{\includegraphics{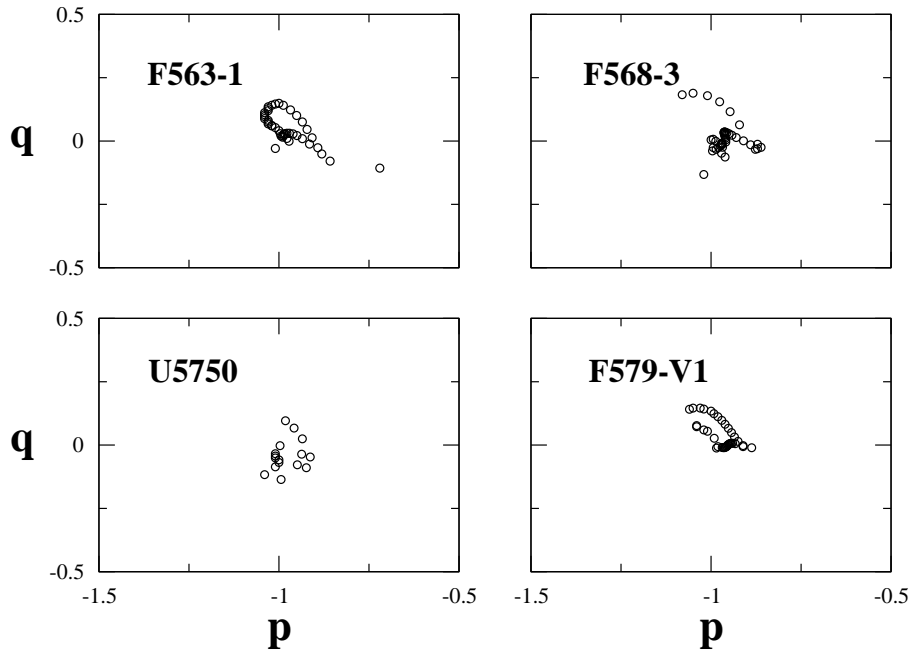}}
\caption{ \label{figE} Numerical estimates of $(p,q)$. Note: Each box is a unit
box in the $(p,q)$ plane.}
\end{center}
\end{figure}
\begin{figure}
\begin{center}
\resizebox{12cm}{!}{\includegraphics{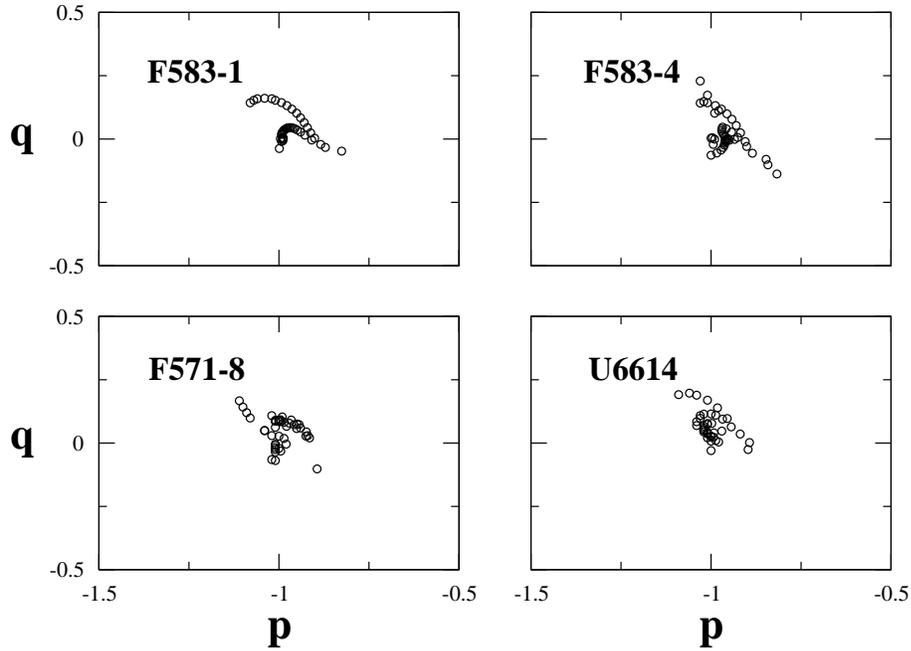}}
\caption{ \label{figF} Numerical estimates of $(p,q)$. Note: Each box is a unit
box in the $(p,q)$ plane.} \end{center}
\end{figure}
\subsection{The Parameter $m_0$}
Referring to (\ref{Mass1}), and noting that $\rho$ in this equation is the
density of material in a disc of unit thickness, we can deduce from the
definition of $\psi$ given there that $m_0$ has dimensions of mass.
This suggests that it is likely to be a simply related to the total mass of the
object being modelled. In fact, we found that, defining $M_{gal}$ as the total
or ordinary mass (stars + dust + gas) in the galaxy being modelled (as estimated
by the original observing astronomer - Stacy McGaugh in the present case), then
$m_0=4M_{gal}$ worked extremely well for all eight cases.
\subsection{Estimating The Parameters $(m_0,p,q)$}
We comment firstly that the theory is indifferent to the signatures of $p$ and
$q$ but, for the sake of being explicit, we take them to be always negative.
With this proviso, then as a means of (a) obtaining estimates of $(m_0,p,q)$ for
each LSB in the sample and (b) providing a preliminary consistency check on the
model equations (\ref{Mass1}) and (\ref{VelEqn}), we proceeded as follows:
\begin{itemize} \item Assume that $m_0$ is some simple multiple (fixed for all
the objects) of estimated total mass (obtained by integrating McGaugh's data for
each object); \item For each LSB, use the McGaugh
mass-distribution estimates and rotation velocity measurements to obtain smooth
cubic spline models of mass and velocity distributions;
\item  Use these smooth cubic spline models to obtain estimates of density,
density gradient, velocity and velocity gradient at a sequence of distinct
points across the disc of each LSB;
\item  With these estimates, the differential equations , (\ref{Mass1}) and
(\ref{VelEqn}), at any given point in a disc become two {\it algebraic}
equations for determining $(p,q)$;
 \item For each LSB, solve this pair of algebraic
equations at several points across the disc, and look for consistency in the
resulting sequence of estimates of $(p,q)$. \end{itemize}
A necessary condition of the theory's consistency is that, for some reasonable
choice of $m_0$, the foregoing process will lead to a consistent set of
$(p,q)$ estimates across each LSB disc.

In fact, we found that setting $m_0 = 4 M_{gal}$, where $M_{gal}$ is the
estimated total of ordinary galaxy mass (stars+dust+gas) for each LSB, gave a
very consistent picture for estimates of $(p,q)$. The big surprise was that, not
only was this the case for each individual object, but that the set of $(p,q)$
pairs for the whole sample lies in the same very small neighbourhood of the
$(p,q)$ plane. Figures \ref{figE} and \ref{figF} show the results of this latter
exercise for each of the eight LSBs in our sample, and we see that, for each
object, the $(p,q)$ estimates all lie in the neighbourhood of $(-1,0)$ in the
$(p,q)$ plane. Note: for each LSB, solutions were sought in a very large $(20
\times 20)$ region of the $(p,q)$ plane. The only solutions found  are those
indicated.

This exercise allowed us to conclude that the model equations were highly
consistent with the phenomenology, and also gave us a good starting estimate of
$(p,q) \approx (-1,0)$ for the detailed modelling process of each object in the
sample.
\begin{table}[h] \caption{\label{Table1}}
\begin{center}
\begin{tabular}{|l|c|c|} \hline
Galaxy & $p$ & $q$ \\
\hline
F563-1 & -0.990 &-0.038 \\
F568-3 & -0.970 & -0.004  \\
U5750 & -0.955 & -0.158  \\
F579-V1 & -0.997 & -0.022 \\
F583-1 & -0.980 & -0.047  \\
F583-4 & -0.950 & -0.084  \\
F571-8 & -0.952 & -0.041  \\
U6614 & -0.995 & -0.032 \\
\hline
\end{tabular}
\end{center}
\end{table}
\subsection{The Detailed Models}
The integrated solutions for the rotational velocities and the mass
distributions (using the $(p,q)$ estimates listed in Table \ref{Table1}),
together with the corresponding observational measurements, are given in Figures
\ref{figA} \& \ref{figB}. In every case, we see that the fit of the computed
rotation velocities (solid lines) to the measured rotation velocities (filled
circles) is, for all practical purposes, perfect.

Except for discrepencies near galactic centres, where our modelling
assumptions become less good,  our computed density distributions (dotted lines)
provide reasonable fits to the estimated densities (crosses). However, it is
important to realize that, by contrast with the accuracy of Doppler velocity
measurements, mass estimation in galaxy discs is subject to very great
uncertainties - which is why mass-modellers never quote error-bars for their
estimates. We have used McGaugh's mass-models to estimate our parameter $m_0$,
mentioned earlier. But, for the mass-density integration, we have chosen the
initial conditions to ensure that the predicted mass distributions over the
discs provide good qualitative fits to MacGaugh's models.

Finally, it should be remarked that all dark matter models fail comprehensively
when applied to LSBs - MOND (described in detail in \S\ref{MOND}) has been
successfully applied to all of the objects considered here (deBlok \& McGaugh
\cite{deBlok1998} and McGaugh \& deBlok \cite{McGaugh1998b})  and to a great
many more.
\begin{figure}
\noindent
\begin{minipage}[b]{\linewidth}
\resizebox{13cm}{!}{\includegraphics{Fig9.eps}}
\end{minipage}
\caption{ Solid line = calculated velocity; dotted line = calculated mass
density; \hfill \break circles = measured velocities with error bars; crosses =
estimated mass density. \hfill \break Arrows = switch-points.} \label{figA} \end{figure}
\begin{figure}
\noindent
\begin{minipage}[b]{\linewidth}
\resizebox{13cm}{!}{\includegraphics{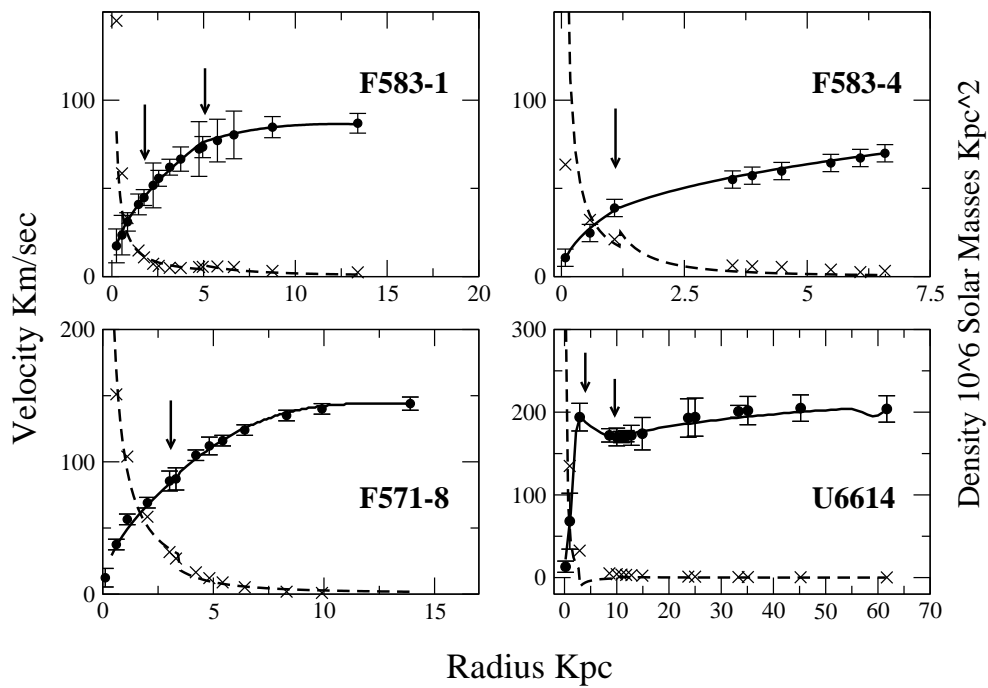}}
\end{minipage}
\caption{Solid line = calculated velocity; dotted line = calculated mass
density; \hfill \break circles = measured velocities with error bars; crosses =
estimated mass density. \hfill \break Arrows = switch-points. }
\label{figB}
\end{figure}
\subsection{Remarkable Circumstantial Evidence}
We have already noted
how, for each LSB in the sample,  $(p,q)\approx (-1,0)$. This circumstance
allows us to discover a remarkable connection between the phenomenology, as
represented by these particular $(p,q)$ values and the theory as represented by
the surfaces $\alpha=F_i(p,q),~i=1..6$ defined at (\ref{AlphaConstraint}).

\begin{figure}
\begin{center}
\noindent
\resizebox{8cm}{!}{\includegraphics{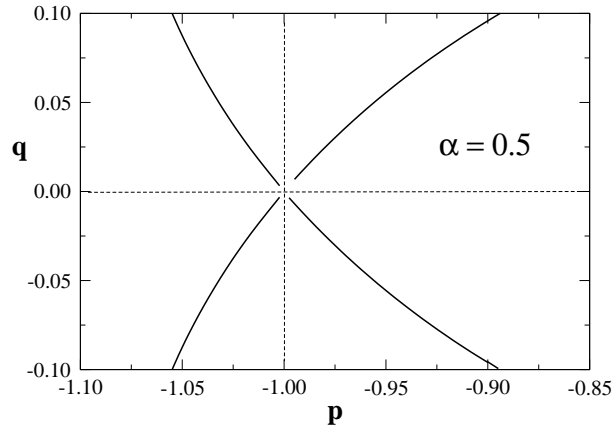}}
\caption{Intersection of  $\alpha=F_i(p,q),~i=1..6$ in the plane
$\alpha=0.5$}
\label{FigC}
\end{center}
\end{figure}

Specifically, we know that, in practice, it is virtually always the case
that $0< \alpha <1$ (see Figure \ref{fig1}, for example) so that we might expect
the intersection of these surfaces with the plane-surface $\alpha=0.5$ to give a
fairly typical cross-section. Figure \ref{FigC} shows this cross-section in the
neighbourhood of $(p,q)=(-1,0)$, which contains our LSB sample. We see
immediately that the point $(p,q)=(-1,0)$ enjoys a very special status in
this $\alpha=0.5$ plane - it is, in fact, the point of intersection of {\it
four} distinct surfaces from the set $\alpha=F_i(p,q),\,i=1..6$ with the plane
$\alpha=0.5$.  Closer investigation reveals that $(p,q)=(-1,0)$ retains its
status as a special nodal point for all $(0<\alpha<1)$, and is therefore a {\it
distinguished axis} for the theory.

So, we have the circumstance that our LSB sample lies in the neighbourhood of a
distinguished axis of the theory defined by the intersection of four particular
surfaces. Whilst it is probably not possible to say what the meaning of
this is at present (we need larger samples of LSBs and corresponding samples of
ordinary spirals with complete mass models), there is a very good chance that it
represents a circumstance of considerable significance in the overall context of
galactic evolution and dynamics.
\begin{figure} \begin{center}
\noindent \resizebox{13cm}{!}{\includegraphics{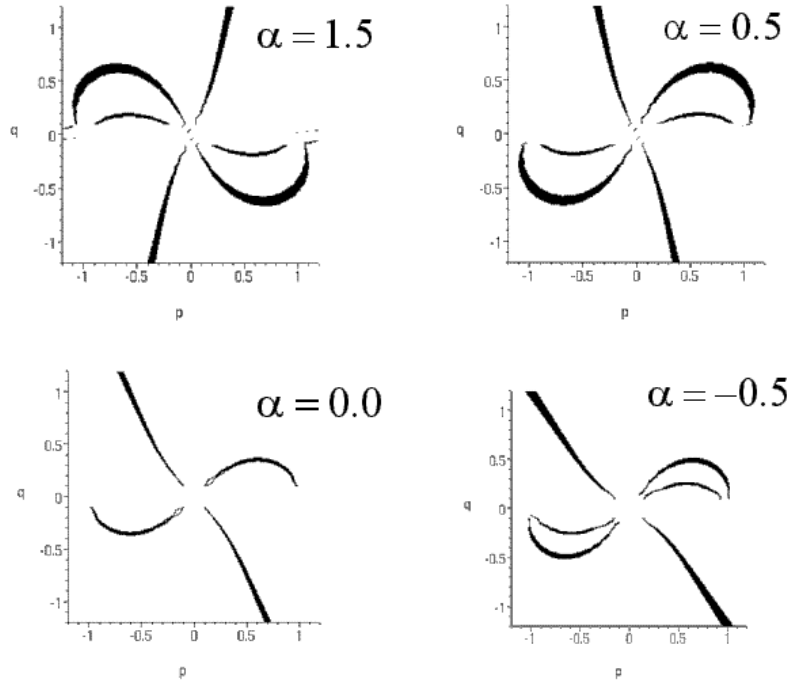}}
\caption{Intersection of  $\alpha=F_i(p,q),~i=1..3$ in the planes
$\alpha=1.5, 0.5, 0.0, -0.5$}
\label{FigG}
\end{center}
\end{figure}
\subsection{Global Complex Structure and $\alpha$ Phenomenology}
\label{GCSTT}
Figure \ref{FigG} shows the evolution of three of the six surfaces as $\alpha$
varies in the range $(1.5,-0.5)$ (the three omitted surfaces are mirror images
of the three shown).  At $\alpha=1$ (not explicitly shown) there is a degeneracy
as the three curves merge identically along the $q=0$ axis whilst, at
$\alpha=0$ (explicitly shown), there is a degeneracy in which two of the curves
merge identically, although not along any axis. Thus, the two values $\alpha=0$
and $\alpha=1$ which, according to figure \ref{fig1}, appear to bound the
phenomenology, are also each associated with degenerate transitions of the
surface topology.
\section{More On The Discrete Dynamical Classes Phenomenology}
\label{MODDCP}
The LSB modelling exercise of \S\ref{LSBs} was significant at two
levels:
\begin{itemize}
\item firstly, it showed how the theory successfully models the details of
LSB dynamics and mass distributions;
\item secondly, it led us to the significant
discovery that LSBs are strongly associated with the distinguished axis
$(p,q)=(-1,0)$ in the $(\alpha,p,q)$ parameter space.
\end{itemize}
The second point poses questions relevant
to gaining a detailed understanding of the discrete dynamical states
phenomenology.
Specifically, according to the discrete dynamical states phenomenology,
{\it low-luminosity} spirals are associated with the $A$-state typified in
figure \ref{fig3}. The LSBs of our sample are, by definition,
low-luminosity spirals and are therefore, presumably, $A$-state spirals. But,
we have also seen that these LSBs are strongly associated with the
distinguished axis, $(p,q)=(-1,0)$, and so the obvious question now is:
what is the distribution of  higher-luminosity spirals ( that is, $B$, $C$ and
$D$-state spirals) in the $(p,q)$ plane?

This question must be answered before we can begin to understand the discrete
dynamical states phenomenology in any detailed way. But, at the moment, we
cannot address this question rigorously, simply because we do not have available
a sample of higher-luminosity spirals with detailed estimated mass distributions
(stars+dust+gas). The reason for this lack of data is simple: since, according
to the canonical view,  $>95\%$ of galaxy mass is CDM then there is no point in
constructing detailed maps of the distribution of ordinary matter in galaxy
discs. This exercise has only been done systematically (and for LSBs only)  by
astronomers interested in the MOND vs CDM debate.

However, there is a possible way forward: for every one  of the $\approx
2500$ ordinary spirals used in the four discrete dynamical states analyses we
have $I$-band (or $R$-band) photometry. In principle, such photometry can be
used to derive broad-brush estimates of ordinary mass distributions which,
combined with large statistics, might be sufficient for the task at hand. This
is for future work.
\section{The Detailed Dynamical Theory}
\label{ISG1}
In
the following, we finally detail the application of the basic theory to the
modelling of an idealized spiral galaxies - defined to be one without a central
bulge, and without the irregularities that are routinely present in the discs of
real galaxies, and without the bars that are present in many spirals. We model
this idealized spiral galaxy as a mass distribution possessing perfect
cylindrical symmetry.

We know, from the considerations of \S\ref{Sec8}, that the general form of the
metric tensor, for any given material distribution, is
\begin{displaymath}
g_{ab} \equiv \nabla_a \nabla_b U
\equiv {\partial^2 {\cal M} \over \partial x^a \partial x^b}
- \Gamma^k_{ab} {\partial {\cal M} \over \partial x^k},
\end{displaymath}
where $\Gamma^k_{ab}$ represents the metrical affinity and where, by the
conjecture of \S\ref{Conjecture}, the value of ${\cal M}({\bf r})$ is
to be generally understood as a measure of the amount of (ordinary) mass
contained within the level surface passing through the point with position
vector ${\bf r}$ defined with respect to the mass-centre of the  ordinary
perturbing matter distribution. In the present case, of course, the level
surfaces are infinite cylinders, and so we have to modify the definition so that
${\cal M}({\bf r})$ refers to the mass contained within cylinders of unit
thickness.  With this understanding, and remembering that ${\cal M}' \equiv
d{\cal M} /d \Phi$ and that $\Phi = R^2/2$, it is easily shown that ${\cal M}'
\equiv 2 \pi \rho$, where $\rho$ is the mass density in a disc of unit
thickness.

Noting that the geometry on any
level-surface of a cylindrical distribution is Euclidean, it follows that the
$\Gamma^k_{ab}$ are identically zero in the present case. Consequently,
\begin{equation} g_{ab} = {\partial^2 {\cal M} \over \partial x^a \partial x^b}
\equiv {\cal M}' \delta_{ab} + {\cal M}'' x^a x^b, \label{eqn06} \end{equation}
where we remember ${\cal M}' \equiv d{\cal M}/ d\Phi$ and $\Phi\equiv R^2/2$
(this latter notation simplies the algebra). Defining the Lagrangian density
\begin{displaymath} {\cal L} \equiv \sqrt{g_{ij} {\dot x}^i {\dot x}^j} = \left(
{\cal M}' <{\dot {\bf R}}|{\dot {\bf R}}> + {\cal M}'' {\dot \Phi}^2
\right)^{1/2}, \end{displaymath}
where ${\dot x^i} \equiv dx^i/dt$ etc., the equations of motion are found
from the Euler-Lagrange equations as
\begin{eqnarray}
&~&2 {\cal M}' {\bf {\ddot R}} +
\left(2 {\cal M}'' {\dot \Phi} - 2{{\dot {\cal L}} \over {\cal L}} {\cal M}'
\right) {\bf {\dot R}} \nonumber \\
&~&+ \left( {\cal M}''' {\dot \Phi}^2 + 2 {\cal M}'' {\ddot \Phi} -
{\cal M}'' <{\bf {\dot R}}|{\bf {\dot R}}>
- 2{{\dot {\cal L}} \over {\cal L}} {\cal M}'' {\dot \Phi} \right) {\bf R} = 0.
\label{eqn1}
\end{eqnarray}
However, it is obvious that the Lagrangian density, defined above, will lead to
a variational principle which is degree zero in the {\lq time'} parameter.
It follows that the equations of Euler-Lagrange pair, above, cannot be
linearly independent.
Whilst either equation can therefore be chosen, it transpires that
the ${\bf \hat \theta}$ component equation is algebraically less complicated.
\subsection{The ${\bf \hat \theta}$ component}
We have
\begin{displaymath}
2 {\cal M}' ( R \,{\ddot \theta}+2\,{\dot R}\,{\dot \theta}) +
R {\dot \theta} \left( 2 {\cal M}'' {\dot \Phi}-2{{\dot {\cal L}} \over {\cal L}} {\cal M}'
\right)
= 0.
\end{displaymath}
Multiply through by $R/ 2{\cal L}$ and use
${\cal M}'' {\dot \Phi} \equiv {\dot {\cal M}'}$ to get, after some
rearrangement:
\begin{displaymath}
\left({{\cal M}' \over {\cal L}}\right)\,{d \over dt} (R^2 {\dot \theta})
+  R^2\,{\dot \theta}\,
\left({{\cal L}\,{\dot {\cal M}'}\,-\,{\dot {\cal L}}\,{\cal M}' \over {\cal L}^2}
\right) = 0\,.
\end{displaymath}
This integrates directly to give:
\begin{displaymath}
{d \over dt}\, \left\{ R^2\,{\dot \theta}\,\right.
\left. \left({{\cal M}' \over {\cal L}} \right) \right\} = 0,
\end{displaymath}
from which we see that angular momentum is \textit{not} generally
conserved. Consequently, the net disc forces are not, in general,
central forces so that, correspondingly, there exists a mechanism
for transferring angular momentum through the disc.
This latter equation integrates to give:
\begin{equation}
({\cal M}')^2\,R^2\,V_\theta^2 =
m_0 \left[ ({\cal M}'+{\cal M}''\,R^2)\,V_R^2+{\cal M}'\,V_\theta^2 \right]
\label{eqn01}
\end{equation}
where $V_R \equiv {\dot R}$ and $V_\theta \equiv R\, {\dot \theta}$.
\subsection{Completion of the Dynamical System}
\label{Sec11}
The cylindrical symmetry of the idealized spiral galaxy implies that there
is net zero force out of the plane of the galaxy.
It follows that the {\lq self-similar'} dynamics condition
(cf \S\ref{SymForces}), which must be used in conjunction with (\ref{eqn01}),
to close the system can be written as:
\begin{equation}
{Transverse~Accn \over Radial~Accn } = k_0.
\label{eqn2a}
\end{equation}
where, because of the radial symmetry of the system, $k_{0}$ has
the same value along all radial directions. To obtain the quantitative
form of this, we need expressions for the radial and transverse accelerations
in the disc geometry. These are derived in appendix \ref{Sec11A},
and we find that (\ref{eqn2a}) becomes
\begin{eqnarray}
&-&k_0\,S\,\sqrt{{ {\cal M}' + {\cal M}''\,R^2 \over {\cal M}' }} \,V_R\,{d V_R \over d R} \,+\,
 V_R \, {d V_\theta \over dR} = \nonumber \\
&-&\left( S\,\sqrt{ {\cal M}' + {\cal M}''\,R^2 \over {\cal M}' }
\,+\, { {\cal M}'' \over 2 \,{\cal M}'} \,R^2\, \right)\,
{{V_R}\,{V_\theta} \over R} \nonumber \\
&+&\,k_0\,{\left( S\,{3 {\cal M}'' + {\cal M}'''\,R^2 \over
2 \,\sqrt{ {\cal M}'({\cal M}' + {\cal M}''\,R^2 )}} \,R\,{V_R}^2
\,-\,
\,{ V_\theta^2 \over R} \right) }  \, \label{eqn02}
\end{eqnarray}
where $S = \pm 1$.

Remembering that ${\cal M}'=2 \pi \rho$ (cf start of \S\ref{ISG1}),  we now note
that  the expression ${\cal M}' + {\cal M}'' {R}^2$ is {\it negative} if  $\rho$
drops off more quickly than $1/R$ - which,  in practice, always seems to be the
case. The implication of this practical reality is that $k_0$ is actually
complex so that the two equations (\ref{eqn01}) and (\ref{eqn02}) represent
three equations in the three unknowns $(V_\theta,V_R,{\cal M}')$, and not just
two. It is when $V_R$ is eliminated between these three equations that we obtain
equations (\ref{Mass1}) and (\ref{VelEqn}) of \S\ref{GSCS}.

However, we initially overlooked the possibility that  ${\cal M}' + {\cal M}''
{R}^2$ might be negative and, consequently, believed we needed a further
equation to close the system completely. As it happens, our approach to
obtaining this extra equation gave ${\cal M}' + {\cal M}'' {R}^2<0$
anyway, and effectively picked out a special solution of the above system, and
is described in the following section. \section{A Special Case Solution: The
Logarithmic Disc} \label{ISG2} Our oversight led us to believe that a further
equation was required to close the system. We argued as follows: It has been
recognized for a very long time that, if the obvious irregularities which exist
in spiral discs are ignored, then the spiral structure of spiral galaxies is
essentially logarithmic. In the context of a classical disc, the  most direct,
way to interpret this phenomenology is to write \begin{equation} {V_\theta \over
V_R} \,=\, K_1 \label{eqn07} \end{equation}
since this implies directly that disc streamlines are logarithmic
spirals.
Substitution of this into  (\ref{eqn01}), and using ${\cal M}' \equiv 2 \pi
\rho$, gives immediately
\begin{equation}
{\cal M}' \equiv  2 \pi \rho  = { k_2 \over R^2} \label{eqn04}
\end{equation}
for some constant $k_2$.
That is, the density of matter in the logarithmic  disc behaves as an inverse
square law. We quickly find that this implies ${\cal M}'+{\cal M}'' R^2<0$, so
that we are back to the point we overlooked initially - that (\ref{eqn02})
represents, in fact, two distinct equations which must both be satisfied. More
particularly, putting  (\ref{eqn07}) and (\ref{eqn04}) into (\ref{eqn02}) we
find
\begin{displaymath}
(K_1 -k_0 Sj  ) \frac{1}{V_R} \frac{dV_R}{dR}=-
\frac{1}{R} ((Sj-1)K_1  + k_0 (S j +K_1^2) ),
\end{displaymath}
where $S=\pm1$ and $j = \sqrt{-1}$.
Since solutions must be real, we must have $k_0$ complex. Putting $k_0=q+pj$,
where $(p,q)$ are real parameters, then gives the two equations
\begin{equation}
\frac{1}{V_R}\frac{dV_R}{dR } = \frac{1}{R}
\left(\frac{K_1-q K_1^2+pS}{pS+K_1}\right) ,~~~
\frac{1}{V_R}\frac{dV_R}{dR } =
\frac{1}{R} \left(\frac{S K_1+p K_1^2+qS}{qS}\right) \equiv
\frac{\alpha}{R} \label{TwoEqns} \end{equation} which must be identical. This
latter requirement quickly gives that the parameters $(K_1,p,q)$ must satisfy
\begin{equation}
p K_1^2 +S(1+p^2+q^2)K_1 +p=0.       \label{AlphaEqn}
\end{equation}
If we use this to eliminate the explicit appearance of $K_1^2$ in the second of
(\ref{TwoEqns}), we get
\begin{displaymath}
\alpha = 1-\frac{p}{Sq} - \frac{p^2+q^2}{q }K_1=1-\frac{p}{Sq} -
\frac{p^2+q^2}{q K_1}. \end{displaymath}
The second expression for $\alpha$, which is the form given at
(\ref{AlphaConstraint}), occurs since, if $K_1$ is a solution of
(\ref{AlphaEqn}), then so is $1/K_1$.

\section{The MOND Programme}
\label{MOND}
\subsection{Overview of MOND}
Modified Newtonian Dynamics (MOND) is an empirically motivated modification
of Newtonian gravitational mechanics which can be interpreted as either a
modification of Newton's gravitational law, or as a modification of the
second law (the law of inertia).
See Milgrom \cite{Milgrom1994}
for a comprehensive review. The basic idea was conceived by Milgrom
\cite{Milgrom1983a}, \cite{Milgrom1983b}, \cite{Milgrom1983c},
as a way of understanding the flat rotation curve phenomenology of spiral
galaxies without recourse to the dark matter idea. The basic hypothesis is 
that, in extremely weak gravitational fields ($g << 10^{-10} m s^{-2}$), the 
nature of Newtonian gravitation changes
in such a way that flat rotation curves are the natural result.

The quantitative idea can be briefly described as follows: if we use 
$g_N({\bf R})$ to denote the gravitational acceleration at any position 
${\bf R}$ in a material distibution {\it according to Newtonian theory}, then
the MOND prescription says that the actual gravitational acceleration is given
by $g = \sqrt{a_0\,g_N({\bf R})}$ when the field is {\it extremely weak}. 
Here $a_0$ is a parameter which has been fixed (Begeman et al \cite{Begeman}) for all
applications of MOND to the value $a_0 \,=\, 1.2 \times 10^{-10}\,m s^{-2}$.
Thus, for a particle in a circular orbit about a
point source of mass $M$, the acceleration-balance equation in the MOND limit
of the very weak field would be given by
\begin{displaymath}
{V_{rot}^2 \over R} = \sqrt{{\gamma M \over R^2}\,a_0}~~\rightarrow~~ V_{rot} =
(\gamma\,M\,a_0)^{1/4}\,,
\end{displaymath}
where $V_{rot}$ is the circular velocity.

This simple idea, which in practice, leads to a model with only one
free parameter (the so-called mass-to-light ratio, $\gamma^{*}\,\equiv\, M/L$,
which gives the conversion factor from the observed light to the inferred
mass), has been remarkably successful in explaining the phenomenology
associated with a very wide variety of circumstances (de Blok \& McGaugh
\cite{deBlok1998}). Furthermore, MOND makes several strong predictions
(Milgrom \cite{Milgrom1983b}) and is therefore eminently falsifiable.
It is notable, therefore, that it never has been although several
attempts have been made (de Blok \& Mcgaugh \cite{deBlok1998}). This
circumstance is to be compared with that of the conventionally favoured
CDM (Cold Dark Matter) model. This latter model is a multi-parameter
model, which makes no predictions - other than that of the existence
of CDM - and can therefore never be falsified in the classical sense.
\subsection{General Comments on MOND and the Present Theory}
If the MOND prescription is an accurate reflection of the reality, then there
necessarily exists an underlying theory of gravitation which provides
flat rotation curves in the weak field regime independently of dark matter
distributions. Conversely, any such theory - should it exist - must
necessarily give rise to the basic MOND prescription, 
$g\,=\,\sqrt{a_0\,g_N({\bf R})}$, in the weak field when interpreted from a
Newtonian perspective.

In the present case, flat rotation curves correspond to the degenerate
$\alpha=0$ case discussed in \S\ref{GCSTT}. Thus,
flat rotation curves solutions do exist in the presented
theory, and are associated with a degeneracy, and are therefore
special in some - yet to be understood - way. The inclusion of the qualitative
aspects of MOND in the theory is already guaranteed by the simple fact that flat
rotation curve solutions are admitted.
\subsection{LSBs - Extreme Objects in MOND and the Present Theory}
According to any given theory, when the mass/dynamical relations in a
given spiral object cross a certain threshold specific to that theory, the
object ceases to be gravitationally bound. In the early 1980s Milgrom realized
that, according to MOND, there should exist spiral objects which were so
diffuse that they could not possibly exist according to the canonical theory.
These objects, now known as Low Surface Brightness galaxies (LSBs), were
subsequently observed in the late 1980s. Of all astrophysical objects, these
present the most critical conditions for the CDM models since,
according to these models, LSBs must typically consist of more than 99\% dark
matter.
Even so, it is now well recognized that the CDM models have suffered
comprehensive failure when applied to model LSBs.

Beyond the initial prediction, the importance of LSBs to MOND can be summarized
as follows: a common criticism of MOND is that, since it was designed to yield
flat rotation curves in the very weak-field regimes of spiral exteriors, it is
hardly surprising that it does so. However, LSBs are so low-mass and diffuse
that virtually the whole of the typical LSB disc is in the MOND regime -
including the rising segments. Thus, the very successful application of MOND to
LSBs has completely undermined such criticisms.

However, we have shown in this paper that a sample of LSBs of widely differing
dynamical properties are modelled virtually perfectly by the presented theory.
We can therefore conclude that the quantitative aspects of MOND must also
be included within it.
In practice, of course, the theory is very much more difficult to
apply than MOND because, unlike MOND which uses the observed mass distributions
to make its dynamical predictions, it requires the mass distributions to be
calculated using (\ref{Mass1}) - and this is a very difficult equation to
integrate because the switching points must (currently) be found by trial and
error.

\section{Conclusions}
The fractal $D=2$ inertial universe (Roscoe \cite{RoscoeC}) provides an entirely new way
of understanding the idea of {\lq inertial space \& time'}, and can be 
considered as the strongest possible realization of Mach's Principle.
We have considered how gravitational processes might arise in such a universe,
and have indicated how to derive the dynamical equations for extended
high-symmetry mass systems. The process has been explicitly illustrated by
applying it to derive the model equations for an idealized spiral galaxy,
defined as one possessing perfect cylindrical symmetry.

The parameter space, $(p,q,\alpha)$, of these dynamical equations has a
complicated topology, and we have shown how various aspects of the phenomenology
have a ready qualitative explanation in terms of it - in particular,
the {\it discrete dynamical states} phenomenology falls into this category.

The theory has been very successfully applied to model
the dynamics and mass distribution of eight Low Surface Brightness spiral
galaxies which, hitherto, have been successfully modelled only by the MOND
algorithm introduced by Milgrom \cite{Milgrom1983a}, \cite{Milgrom1983b},
\cite{Milgrom1983c}. The CDM models inevitably fail badly in this context.  Of
equal significance to the theory's success in this context is the fact that the
values necessarily assigned to the parameters $(p,q)$ for each of the eight LSBs
are all in the neighbourhood of $(p,q)=(-1,0)$ - which, it transpires, is a {\it
very special} distinguished axis in the $(p,q,\alpha)$ parameter space.  As well
as providing further circumstantial support for the theory, this latter fact
suggests the possibility of intimate connections between the topology of the
parameter space and galactic evolution.

To summarize, we have presented a theory with a very richly structured parameter
space and have shown how several aspects of spiral galaxy phenomenology fit
beautifully into this structure.  There would appear to be every prospect that
the theory can form the basis of understanding spiral galaxies and their
evolution in a way that has hitherto not been possible.  Only very much more
work involving the analysis of many more spirals of all types will show if this
bold claim can become a reality.

\appendix
\section{Preliminaries}
\label{Prelim}
Mass in the  $D=2$  equilibrium universe of Roscoe \cite{RoscoeC}
is distributed according to
\[
M = \frac{m_0}{r_0^2} r^2,
\]
so that $m_0$ is the amount of mass contained in side a sphere of arbitrary
radius $r_0$, and $g_0 \equiv m_0/r_0^2$ is a global constant of this
equilibrium universe.

However, this model universe  is, in fact, a particular case of a class of
non-equilibrium model universes possessing a general spherical symmetry. But,
since we are primarily interested in non-spherical systems, we use the $D=2$
equilibrium solution as our starting point, and consider how to perturb that in
progressively complex ways.

The general spherically symmetric model has an associated potential function
defined in Roscoe \cite{RoscoeC} as
\begin{equation}
V \equiv
C_0 - \frac{v_0^2}{4\,d_0^2 \,g_0} A + {B \over 2A}r^2{\dot r}^2,~~
\label{(1)} \end{equation}
where $C_0$ is the arbitrary constant usually associated with potential
functions, $v_0$ is a global constant having the dimensions of velocity,
$g_0\equiv m_0/r_0^2$  is the global constant  defined above, $d_0$ is a
dimensionless constant evaluated below, and
\begin{equation} A \equiv { 2{\cal
M} \over r^2},~~~ B \equiv -\left( {2{\cal M} \over r^4} \right) +
{ {\cal M}' {\cal M}' \over 2 d_0 {\cal M}},~~~
{\cal M}' \equiv \frac{d {\cal M}}{d \Phi},~~~\Phi\equiv \frac{r^2}{2}
\label{(2)} \end{equation}
where  the notation $\Phi\equiv r^2/2$ is introduced to simplify the algebra
later on, and \begin{equation}
{\cal M}=d_0 M+m_1,~~~~~
M = m_0\left(\frac{r}{r_0}\right)^2+2\sqrt{\frac{m_0
m_1}{d_0}}\left(\frac{r}{r_0}\right).\label{(2A)} \end{equation}
As before, $M$ quantifies the mount of mass
inside a sphere of radius $r$ whilst $m_1$ an arbitrary constant having units
of mass which quantifies the
perturbation from the $D=2$ equilibrium universe. The
dimensionless constant $d_0$ can be determined by noting
that the special case $m_1=0$  must recover the equilibrium (inertial) case,
which requires $A= const$ and $B=0$. Reference to the above shows that this can
only happen if $d_0 = 1$, and so this value is assumed from hereon.
\subsection{Interpretation Issues} The equilibrium case
($V=const$) of globally inertial conditions arises when $d_0 = 1$, $m_1 = 0$ and
${\cal M}(r) =m_0 r^2/r_0^2$ about any centre. Since ${\cal M}(r)$ (in the
form of $M$) is interpreted as the amount of mass in a sphere of radius $r$,
then it follows immediately that globally inertial conditions are irreducibly
associated with a fractal $D=2$ mass distribution. However, it was noted in
Roscoe \cite{RoscoeC} that this mass exists in the form of a {\lq quasi-photon'}
gas - that is, it consists of primitive particles moving in arbitrary directions
but in otherwise identical states of motion, in direct analogy with photons in a
vacuum. For this reason, we interpreted it as a {\lq quasi-classical'} vacuum
gas and noted that, assuming conventional material {\lq condenses'} out of this
vacuum gas in some way - perhaps by collision processes - then the theory
provided a direct way of understanding the observed $D \approx 2$ distribution
of galaxies on medium scales.

In the present case, we consider perturbations of ${\cal M}$ that are not
generally spherically symmetric. This has the direct result that our
interpretation of ${\cal M}$ must necessarily evolve such that our original
understanding is included as a special case.
\section{Newtonian Gravitation For Test Particles}
\label{Sec2}
In this section, we establish the principle that
classical Newtonian gravitation can be recovered as a point-mass perturbation
of the $D=2$ equilibrium universe,
${\cal M} = m_0 r^2/r_0^2$ where we remember $g_0 \equiv m_0/r_0^2$  is a
global constant.
But, as a by-product of this analysis, we also establish that the point-mass
perturbation necessarily picks up its conventional mass properties via a global
interaction - as required by conventional interpretations of Mach's Principle.

We consider the most simple possible perturbation of the $D=2$ distribution
(in fact, the original one given in Roscoe \cite{RoscoeC})which (\ref{(2A)})
(with $d_0=1$) shows is given by
\begin{equation} {\cal M} =  \left(
\frac{\sqrt{m_0}}{r_0}\,r+\sqrt{m_1}\,\right)^2  \label{(3)} \end{equation}
where the coordinate origin, ${\bf r}=0$, is the position of the perturbing
mass.

The general form of the potential function, $V$, generated by
an arbitrary spherically symmetric mass distribution, ${\cal M}$, is given by
(\ref{(1)}) with (\ref{(2)}); consequently, to consider the circumstances under
which (\ref{(1)}) gives
rise to Newtonian gravitation - if at all - it is only necessary to consider
the structure of this potential when ${\cal M}$ is defined by (\ref{(3)}).
However, this form of the potential function is an explicit function of
$r$ and ${\dot r}$, which makes analysis more difficult. A more convenient form,
expressed purely in terms of $r$, is given in \cite{RoscoeC}. For the particular
case $d_0=1$, this is given by:
\begin{equation}
 V(r) ~~\,=\,
-{2 v_0^2 \over r} \sqrt{{ m_1 \over  g_0}}
- { 1 \over r^3 } \sqrt{ { m_1 \over g_0} }
\left( {2 m_1 v_0^2 \over  g_0} - h^2 \right)
- {1 \over 2 r^4} { m_1 \over g_0}
\left( { m_1 v_0^2 \over  g_0}  - h^2 \right)
\label{(4)}
\end{equation}
where $h$ is the classical angular momentum.
This can only become a {\it first order} approximation
for potential of classical Newtonian theory if
\begin{equation}
2v_0^2\, \sqrt{\frac{m_1}{g_0}} =\gamma\,M_S \label{(4A)}\end{equation}
where $M_S$ is the conventional mass of the central disturbing distribution,
$\gamma$ is the usual gravitational constant and we remember that
$m_1$, which has units of mass, is a quantitative measure of the perturbing
disturbance in the equilibrium universe.
This relation is extremely interesting since, if $\gamma$ is a
global constant as we believe, then it effectively states that
\[
M_S = \sqrt{M_0 m_1}
\]
where $M_0$ is a global scaling constant with dimensions of mass. In other
words, the perturbing  disturbance, quantified by the parameter $m_1$
originally, picks up its conventional mass properties via a global interaction -
which is a common interpretation of Mach's Principle.

In  terms of (\ref{(4A)}), then (\ref{(4)})
becomes
\[
 V(r) ~~\,=\, -\frac{\gamma M_S}{r}
- \frac{\gamma M_S}{r^3}
\left[ \left(\frac{\gamma M_S}{2v_0^2}\right)^2 - \frac{h^2}{2
v_0^2} \right] - {\gamma M_S \over  r^4}
\left[ \left(\frac{\gamma M_S}{2v_0^2}\right)^2  - \frac{h^2}{v_0^2} \right]
\frac{\gamma M_S}{8 v_0^2}.
\]
It is clear from this expression that standard Newtonian results will be
reproduced, provided $v_0^2$ is sufficiently large.
\section{The Two-Body Problem}
\label{Sec3}
We know from the Newtonian analysis that the two-body problem is essentially
spherically symmetric. Consequently, the two-body mass function ${\cal M}$
necessarily has a structure similar to the one-body case, given at (\ref{(3)}),
except that an extra degree of freedom must be accounted for. This can be most
easily accomplished by a mass-function similar to:
\begin{equation}
{\cal M} =
m_0\left(\frac{r}{r_0}\right)^2 + m_1\left(\frac{r}{r_0}\right)+m_2  \label{(7)}
\end{equation}
where $(m_1,m_2)$ have the dimensions of mass, and will be functions of the
masses of the perturbing sources.

A similar analysis to that of the one-particle case produces a relation similar
to (\ref{(4A)}) - with the major difference that $M_S$ is replaced by the
Newtonian expression for the effective gravitational mass at the mass-centre of
a two-body system.
\section{Global Momentum Conservation And Consequences For ${\cal M}$}
\label{GMC}
Having established how classical Newtonian gravitation can, in principle, arise
as a point-mass perturbation of the inertial fractal $D=2$ universe, we can
consider the question of point-mass perturbations by an ensemble of $N$
conventional particle masses.
But this automatically raises the question of momentum conservation in the
ensemble, which we consider here.

We find the remarkable result that momentum conservation in the finite ensemble
of particles requires that the distribution of vacuum mass, ${\cal M}$, must be
expressible as an {\it even} function of $m {\bf r}$, where $m$ is the
inertial mass of an arbitrarily chosen ensemble particle, and ${\bf r}$ is its
position defined with respect to the ensemble mass centre.
This then leads to a {\it tentative} reinterpretation of ${\cal M}$ as a measure
of the total vacuum mass detected by a particle of inertial mass $m$ at position
${\bf r}$ - or, to be more specific, it suggests that ${\cal M}(m{\bf r})$ is
a measure of the total vacuum mass contained with the level surface which passes
through the point ${\bf R} \equiv |m|{\bf r}$.

However, since we are also
accustomed to thinking that, within a gravitating particle ensemble, the orbit
of a particular mass $m$ is a function of the distribution of the other masses
within the ensemble, then it would appear that the  measure of total vacuum
mass, ${\cal M}(m{\bf r})$, within the level surface {\it must also} be a
measure of the total inertial mass of the finite particle ensemble. The
implication is that there is a deep association between the inertial masses of
the particles in the ensemble and the vacuum mass.

We adopt the tenative working interpretation that, for the case $m=1$, then
${\cal M}({\bf r})$ is a measure of the inertial mass contained within the level
surface which passes through the point which has position vector ${\bf r}$ with
respect to the ensemble mass centre.
\subsection{The Details} Firstly, since we have a finite ensemble embedded in an
equilibrium background, we can suppose that all discussion of momentum
conservation can be referred to the mass centre of the ensemble, and that this
mass centre is in dynamic equilibrium with the background.

For any system of particles of masses $M_1,...,M_N$, described from a
centre-of-mass frame, the integrated momentum-conservation equation becomes
\begin{displaymath}
M_1 {\bf R}^1 + M_2 {\bf R}^2 +...+ M_N {\bf R}^N = 0.
\end{displaymath}
The masses appearing in this equation are now arbitrarily partitioned into the
pair of ensembles $M_1,...,M_{k-1}$ and $M_k,...,M_N$.
Defining the mass of the whole system as $M$, and the mass of the ensemble
$M_1,...,M_{k-1}$ as $m$, then the foregoing equation can be written as
\begin{displaymath}
m {\bf r} + (M-m) {\bf R} = 0,
\end{displaymath}
where ${\bf r}$ and ${\bf R}$ are the respective mass-centres of the
two, arbitrarily defined, particle ensembles defined with respect to the mass
centre of the whole ensemble.
Any interaction can then be considered as being between the particle
ensemble of mass $m$ (which can represent a planet or star or galaxy,
etc) and the rest of the ensemble, having mass ${M-m}$.
Whatever the details of this interaction, these two particle ensembles
must, together, evolve from their initial state in such a way that linear
momentum is conserved for all $t>0$ so that, always,
\begin{equation}
m {\bf r} = - (M-m) {\bf R}.
\label{GMC1}
\end{equation}
Now, we know from classical Newtonian theory that the two-body problem
can be reduced to spherically symmetric form, and so we can expect
the same here. Consequently, from the point of view of either of the
two bodies, perturbations of the mass function, ${\mathcal{M}}$,
are also spherically symmetric about the ensemble mass centre. But,
it is easily shown that the equations of motion for the spherically
symmetric system (Roscoe \cite{RoscoeC}) are scale-invariant under
${\bf{r}}=\lambda{\bf{R}}$ for non-zero constant $\lambda$,
up to unspecified ${\mathcal{M}}$. Consequently, under (\ref{GMC1}),
the structure of the equations of motion remains unchanged up to ${\mathcal{M}}$
not specified. It follows that if ${\mathcal{M}}$ is an \textit{even}
function of $m{\bf{r}}$ then the equations of motion for $m{\bf{r}}$
will transform into \textit{identical} equations of motion for $(M-m){\bf{R}}$
under (\ref{GMC1}) so that, with the initial condition $m{\bf{r}}(0)=-(M-m){\bf{R}}(0)$,
the calculated trajectories will satisfy $m{\bf{r}}=-(M-m){\bf{R}}$
for all time. It is easily seen that no other form of ${\mathcal{M}}$
has this property. It follows that, for global momentum conservation,
${\mathcal{M}}$ must be an even function of $m{\bf{r}}$ - as
stated.
\section{The Three-Body Problem}
\label{TBP}
We now consider the structure of ${\cal M}$ appropriate to the non-spherical
case.
A {\lq degrees of freedom'} argument clarifies the situation - note that we
make the fundamental assumption that individual particle masses in the
perturbing ensemble have no intrinsic angular momentum.

Firstly, consider a two-body system: notionally, in the absence of intrinsic
angular momentum, each particle mass in the system has six degrees of freedom
consisting of three positional and three kinematic.
However, in this simple system, once these are set for one mass, momentum
conservation fixes everything for the second mass and so there are only six
degrees of freedom in total for this case.
But the equations of motion for the {\lq free'} body in this system
are second order in three components, and therefore require all six of
these degrees of freedom for their closure.
It follows that ${\cal M}$ need contain no freedom to describe any positional
or kinematic qualities of the detected mass distribution it describes, and this
is reflected in the simple structure of (\ref{(7)}).

Now consider what happens when a third perturbing mass is introduced: If $m_1$
is any one of the three masses in the system, then ${\cal M}({m_1 {\bf r}})$
quantifies the total effective gravitational mass detected by $m_1$ at
position ${\bf r}$, and
provides the basis for the equations of motion of $m_1$,
and these equations of motion still require only six degrees of freedom for
their closure.
However, the introduction of the third perturbing mass has introduced six extra
degrees of freedom into the system which, since these degrees of freedom are
not required by the equations of motion for $m_1$\,, must therefore be
incorporated into the mass function ${\cal M}$ itself.

Introducing the notation
$r \equiv <{\bf r}|{\bf r}>^{1/2} \equiv <{\textbf r}|{\bf
I}|{\textbf r}>^{1/2}$, where ${\bf I}$ is the unit matrix, and bearing in mind 
the idea that, at large distances from the mass centre, we might expect ${\cal
M}(m {\bf r})$ for any finite ensemble to behave like a point-perturbation, then
the most obvious perturbation of (\ref{(7)}) having the structure ${\cal
M}(m{\bf r})$, and which incorporates the required six degrees of freedom, is
given by \begin{equation} {\cal M} = \kappa_0 m^2 r^2 + \kappa_1 m <{\bf r}|{\bf
I}|{\bf r}>^{1/2} + \kappa_2 + \kappa_3 m^{-1}  <{\bf r}|{\bf A}|{\bf
r}>^{-1/2},~~ \label{(9)} \end{equation} where $m$ is the mass of the chosen
particle and ${\bf A}$ is a positive (semi) definite $3 \times 3$ matrix which
provides the required extra degrees of freedom. The restriction of ${\bf A}$ to
positive (semi) definiteness is imposed by the square-root operation and this,
in turn, is what limits ${\bf A}$ to contain only six free parameters.
\section{The $N$-Body Problem} \label{NBP} The generalization to $N$ bodies
(each one requiring six degrees of freedom) is now fairly obvious, and is given
by \begin{equation}
{\cal M}(m {\bf r}) = \kappa_0 m^2 r^2 +
\sum^{N-2}_{j=1} \kappa_j m^{2-j} <{\bf r}|{\bf A}_k|{\bf r}>^{1-j/2},
~~N>2,
\label{(10)}
\end{equation}
where $\kappa_j,~(j=1,...,N-2)$ are constants which are independent of the
chosen mass, and ${\bf A}_j,~(j=1,...,N-2,~N>2)$ is a
class of positive (semi) definite matrices and ${\bf A}_1 \equiv {\bf I}$.

The equations of motion for $m$ are defined in terms of this ${\cal M}$, and
there will be similar definitions for each of the other masses in the system.
Since the mass detected by any chosen mass is the whole system sans
the chosen mass itself, it follows that, in the most general situation, each
individual particle mass in the perturbing ensemble will have its own unique
set of ${\bf A}_k$ matrices.
However, situations of high symmetry can be imagined in which every mass in
the ensemble will detect identical things; it can also be expected that
circumstances will
exist for which ${\bf A}_1 = {\bf A}_2 = ...= {\bf A}_{N-2}$ for any chosen
particle mass; in such a case, ${\cal M}$ will attain a maximal simplicity for
the chosen particle mass.
 \subsection{The Time-Invariant Subset}
\label{Sec8} It remains to define the equations of motion which correspond to
the mass function defined at (\ref{(10)}). To this end, we note that the
arguments of Roscoe \cite{RoscoeC} which lead to the definition of the metric
tensor, $g_{ab}$, from the mass function, ${\cal M}$, are strictly independent
of any assumptions of spherical symmetry. Consequently, we still have
\begin{equation} g_{ab} \equiv \nabla_a \nabla_b {\cal M} \equiv {\partial^2
{\cal M} \over \partial x^a \partial x^b} - \Gamma^k_{ab} {\partial {\cal M}
\over \partial x^k}, \label{(11)} \end{equation} where  $\Gamma^k_{ab}$ is
determined by the metric affinity. This choice was made because it guarantees
that appropriate generalizations of the divergence theorems exist - which is
necessary if we are to have conservation laws.

Assuming this system implies a unique determination of $g_{ab}$ in terms of
${\cal M}$ (as it
does for the case of arbitrary spherical symmetry), then the
Lagrangian density ${\cal L} = \sqrt{ g_{ij} {\dot x}^i {\dot x}^j }$
can defined, and the equations of motion defined as the
Euler-Lagrange equations.
However, as pointed out in Roscoe \cite{RoscoeC}, since the variational principle
arising from this ${\cal L}$ is invariant with respect to arbitrary
transformations of the {\lq time'} parameter, then physical time has yet to
be defined.
\subsection{Physical Time Defined By The Self-Similarity Of Forces.}
\label{SymForces}
This problem of {\lq physical time'} was resolved in Roscoe \cite{RoscoeC}
by the application of the
condition that {\it all accelerations are directed through the global
mass-centre} - but this condition arises partly from the circumstance of
spherical symmetry, and so is
not appropriate to the general case being considered here.
However, as indicated in Roscoe \cite{RoscoeC}, this latter condition actually
represents an integrated form of the fundamental Newtonian condition:
\hfill \break
\hfill \break
{\bf C0} {\it The action between any two material particles is along the
shortest path (a straight line, classically) joining them.}
\hfill \break
\hfill \break
Consequently, it is this part of the physics which has to be given appropriate
expression in the present circumstance, and which will complete the equations
of motion.

The first relevant point in this connection is the recognition that the
incomplete equations of motion are scale invariant up to the specification
of the mass function ${\cal M}$.
The second relevant point is the recognition that the statement {\bf C0}
above is also a scale-invariant statement.
It follows that when the equations of motion are completed by an appropriate
application of {\bf C0}, they will remain scale-invariant up to the
specification of the mass function.

Let us therefore consider the hypothetically completed equations of
motion for the chosen mass $m$, and suppose that they are represented
as
\begin{equation}
{\bf F} ({\bf r}, {\bf {\dot r}}, {\bf {\ddot r}},{\cal M}(m{\bf r})) = 0,
\label{(12)}
\end{equation}
so the trajectory of $m$ is represented by ${\bf r}(t)$.
Since ${\bf F}$ must be scale-invariant up to the specification of ${\cal M}$
then, under the change of scale
${\bf r} = \lambda {\bf R}$, (\ref{(12)}) becomes
\begin{equation}
{\bf F} ({\bf R}, {\bf {\dot R}},{\bf {\ddot R}},{\cal M}( \lambda m {\bf R})) = 0,
\label{(13)}
\end{equation}
which can be interpreted to mean that the trajectory of a chosen mass
$\lambda m$ is given by ${\bf R}(t)$.

Now consider a very special case: since $m{\textbf{r}}(t)$ in (\ref{(12)})
satisfies the same equation as $\lambda m{\textbf{R}}(t)$ in (\ref{(13)})
then, for properly matched initial conditions,
$m{\textbf{r}}(t)=\lambda m{\textbf{R}}(t),\, t\geq0$.
That is, under these specially chosen initial conditions, the trajectory
${\textbf{R}}(t)$ of mass $\lambda m$ is geometrically similar to
the trajectory ${\textbf{r}}(t)$ of mass $m$.

But we now note that a sufficient \textit{dynamical} condition for
the trajectories to be geometrically similar in this way (given the
special initial conditions) is that the force system acting on mass
$\lambda m$ is geometrically similar to the force system acting on
mass $m$. That is, if the resolved force components acting on the
chosen mass are denoted by $(F_{1},F_{2},F_{3})$, then along any
radial drawn from the system mass-centre we have
\begin{equation}
\frac{F_{2}}{F_{1}}=k_{0},~~\frac{F_{3}}{F_{1}}=k_{1}\label{Force_Law}
\end{equation}
for parameters $k_{0},\, k_{1}$ which are constant along any given
radial direction. Of course, in cases of perfect radial symmetry $(k_{0},\, k_{1})$
will be global constants. Since, generally speaking, we do not expect
the dynamical constraints acting in a system to depend on the initial
conditions, then we can take (\ref{Force_Law}) to be a general dynamical
law in the system.

Finally, we note that, since (\ref{Force_Law}) can only be deduced
from the completed dynamical system (\ref{(12)}), then the \textit{incomplete}
dynamical system can be completed by augmenting it with (\ref{Force_Law}).
In other words, the application of (\ref{Force_Law}) to the system
is equivalent to applying the Newtonian law \textbf{C0}.

\section{The Level-Surface Geometry For Very Large $N$-Body Systems}
\label{Sec9}
The mass function for a chosen particle, mass $m$, in a system of $N$ massive
particles is defined at (\ref{(10)}).
For $N=0$ (the inertial fractal $D=$ universe), the level surfaces of ${\cal M}$
are spherical surfaces centred anywhere. For $N=1$ they are spherical
surfaces centred on the single perturbing mass. For $N=2$ they are spherical
surfaces centred on the mass-centre of the two perturbing masses.

The addition of perturbing masses beyond $N=2$ made it necessary to introduce
successive ellipsoidal perturbations of ${\cal M}$ so that its level-surfaces
will be successively perturbed to form a sequence of differing level-surface
structures.
It is obvious that a {\it closed} description of the geometry of ${\cal M}$
(and hence of an analytic determination of these level surfaces) will be
difficult (if not impossible) to calculate even for $N=3$, and almost certainly
impossible for any $N$ which is moderately greater than this value.

However, there exists an alternative means of determining - at least
qualitatively - the level-surface geometry of ${\cal M}$ for high-symmetry very
large $N$ systems.
Whatever these level-surfaces represent (for example, surfaces of constant
$g$-magnitude, as in the spherically symmetric case), they must be determined
by the bulk spatial and kinematic properties of the distributions concerned,
and must therefore reflect any symmetries possessed by these properties.
Nature tells us that many large $N$ distributions (in the form of galaxies)
naturally form such structures as spiral discs, ellipsoids, rings etc which
possess various forms of high-symmetry.
For example, in spiral galaxies, the spatial and kinematic information says
{\lq rotating disc'} and so we can reasonably deduce that the level-surfaces of
any corresponding ${\cal M}$ must likewise possess plane rotational symmetry.
Similarly, the rotational symmetry of an elliptical galaxy about its long axis
and its reflective symmetry about the plane through the mass centre which
contains the two minor axes, provides corresponding information about the
internal symmetries of its ${\cal M}$ function.

Once a geometry for the level-surfaces in a given distribution has been
identified (at least qualitatively), then these level-surfaces can be used
to parametrize ${\cal M}$ - exactly as was done for the spherical analysis of
Roscoe \cite{RoscoeC}.
For simple geometries such as spheres, discs and ellipsoids, this
parametrization allows (\ref{(11)}) to be solved for the explicit form of the
metric
tensor, $g_{ab}$, in terms of ${\cal M}$ and its derivatives with respect to
the parametrization.
The equations of motion can then be determined in the same form, and the system
completed by the {\lq similarity of forces'} condition developed in
\S\ref{SymForces}.

\section{Completion of the Dynamical System: Details}
\label{Sec11A}
We require a quantitative expression for (\ref{eqn2a}) in the particular
geometry of our disc.
We begin by noting that, from (\ref{eqn06}), we have
\begin{eqnarray}
ds^2 &=& {\cal M}' dx^i dx^j \delta_{ij} + {\cal M}'' x^i x^j dx^i dx^j 
\nonumber \\
&=& {\cal M}' \left( dR^2 + R^2 d \theta^2 \right)\,+\, {\cal M}'' R^2 dR^2
\nonumber \\
&=& \left( {\cal M}' + R^2 {\cal M}'' \right) dR^2 + {\cal M}' R^2 d\theta^2
\nonumber \\
&=& \left\{ S_0\,\sqrt{{\cal M}' + R^2 {\cal M}''} \,dR\, {\bf { \hat{ R}}}
\,+\,S_1\, \sqrt{{\cal M}'}\,R\,d\theta\,{\bf {\hat {\theta}}} \right\}^2\,.
\nonumber
\end{eqnarray}
where we have introduced the orthogonal unit vectors, ${\bf { \hat{ R}}}$ and
${\bf {\hat {\theta}}}$, and
where $S_0 = \pm 1$ and $S_1 = \pm 1$ and are independent.
Consequently,
\begin{displaymath}
d{\bf s} = \left\{ S_0\,\sqrt{{\cal M}' + R^2 {\cal M}''} \,dR\, {\bf { \hat{ R}}}
\,+\,S_1\, \sqrt{{\cal M}'}\,R\,d\theta\,{\bf {\hat {\theta}}} \right\}\,,
\end{displaymath}
so that the vector velocity is given as:
\begin{equation}
{\bf V} = \left\{ S_0\,\sqrt{{\cal M}' + R^2 {\cal M}''} \,{\dot R}\, 
{\bf { \hat{ R}}}
\,+\,S_1\, \sqrt{{\cal M}'}\,R\,{\dot \theta}\,{\bf {\hat {\theta}}} \right\}.
\label{eqn03}
\end{equation}
From this,we can now calculate the vector acceleration as:
\begin{eqnarray}
{\dot {\bf V}} &=& \left\{ S_0\,{3 {\cal M}'' + 2 \Phi {\cal M}''' \over 
2 \sqrt{ {\cal M}' + 2 \Phi {\cal M}'' }} \,R\,{\dot R}^2 \,+\,S_0\,
\sqrt{ {\cal M}' + 2 \Phi {\cal M}'' } \,{\ddot R} \,-\,S_1\,{\sqrt{{\cal M}'}
\,R\,{\dot \theta}^2 } \right\}\, {\bf \hat R} \nonumber \\
&+&\,
\left\{ S_0 \, \sqrt{ {\cal M}' + 2 \Phi {\cal M}'' }\,{\dot R}\,{\dot \theta} 
\,+\, S_1\,{ {\cal M}'' \over 2 \sqrt{{\cal M}'}} \,R^2\, {\dot R} \,{\dot \theta}
\,+\,S_1\,\sqrt{{\cal M}'} \left( {\dot R}\,{\dot \theta}\,+\,R\,{\ddot \theta}
\right)\right\}\,{\bf \hat \theta}.
\nonumber
\end{eqnarray}
Hence, using $2 \Phi \equiv R^2$, the condition (\ref{eqn2a}) becomes:
\begin{eqnarray}
&\,&{ \left\{  \sqrt{ {\cal M}' + {\cal M}''\,R^2 }\,{\dot R}\,{\dot \theta} 
\,+\, S\,{ {\cal M}'' \over 2 \sqrt{{\cal M}'}} \,R^2\, {\dot R} \,{\dot \theta}
\,+\,S\,\sqrt{{\cal M}'} \left( {\dot R}\,{\dot \theta}\,+\,R\,{\ddot \theta}
\right)\right\} } \, \nonumber \\
&=&\,k_0\,{\left\{ {3 {\cal M}'' + {\cal M}'''\,R^2 \over 
2 \sqrt{ {\cal M}' + {\cal M}''\,R^2 }} \,R\,{\dot R}^2 +
\sqrt{ {\cal M}' + {\cal M}''\,R^2 } \,{\ddot R} \,-\,S\,{\sqrt{{\cal M}'} 
\,R\,{\dot \theta}^2 } \right\} }  \, \nonumber
\end{eqnarray}
where $S = \pm 1$. This can be rearranged as:
\begin{eqnarray}
&\,&{  \,S\,\sqrt{{\cal M}'} \,\,R\,{\ddot \theta}} \, 
-\,k_0\,\sqrt{ {\cal M}' + {\cal M}''\,R^2 } \,{\ddot R}
\nonumber \\
= &-&\,\left[ \sqrt{ {\cal M}' + {\cal M}''\,R^2 }
\,+\, S\,{ {\cal M}'' \over 2 \sqrt{{\cal M}'}} \,R^2
\,+\,S\,\sqrt{{\cal M}'} \, \right]\,{\dot R}\,{\dot \theta} \nonumber \\
&+&
\,k_0\,{\left\{ {3 {\cal M}'' + {\cal M}'''\,R^2 \over 
2 \sqrt{ {\cal M}' + {\cal M}''\,R^2 }} \,R\,{\dot R}^2 
\,-\,S\,{\sqrt{{\cal M}'} 
\,R\,{\dot \theta}^2 } \right\} }  \,. \nonumber
\end{eqnarray}
Using the identities
\begin{displaymath}
{\ddot R } \,\equiv \, {\dot R}\, {d {\dot R} \over dR}\,,~~~~
{\ddot \theta} \,\equiv \,
{{\dot R} \over R} \,{d ( R {\dot \theta}) \over dR}\,-\,
{{\dot R} \over R^2}\,(R\,{\dot \theta) }
\end{displaymath}
together with $V_R \equiv {\dot R}$ and $V_\theta \equiv R\,{\dot \theta}$,
then this last equation becomes:
\begin{eqnarray}  
&-&k_0\,S\,\sqrt{{ {\cal M}' + {\cal M}''\,R^2 \over {\cal M}' }} \,V_R\,{d V_R \over d R} \,+\,
 V_R \, {d V_\theta \over dR} \,= \,
\,-\,\left( S\,\sqrt{ {\cal M}' + {\cal M}''\,R^2 \over {\cal M}' } 
\,+\, { {\cal M}'' \over 2 \,{\cal M}'} \,R^2\, \right)\,
{{V_R}\,{V_\theta} \over R} \nonumber \\
&+&\,k_0\,{\left( S\,{3 {\cal M}'' + {\cal M}'''\,R^2 \over 
2 \,\sqrt{ ({\cal M}')^2 + {\cal M}'\,{\cal M}''\,R^2 }} \,R\,{V_R}^2 
\,-\, 
\,{ V_\theta^2 \over R} \right) }  \,. \label{eqn02A}
\end{eqnarray}

\end{document}